\documentclass[12pt]{article}
\usepackage{latexsym,graphicx,amsmath,amssymb,mathtools}
\usepackage{stackrel}
\usepackage{lscape}
\usepackage{multirow}
\usepackage{hyperref}

\def\highertp{\topmargin=-2.5cm
            \textheight=24cm} 
\def\higher{\topmargin=-.2cm
            \textheight=21cm} 

\def\tck{\hspace{.6pt}\cdot\hspace{1.2pt}}   
\def\til#1{\mbox{\boldmath$#1$}}
\def\gr#1{\boldsymbol{#1}}
\def\lecture#1{\section*{#1}                 
                \addtocounter{section}{1}
                \setcounter{equation}{0}
                \setcounter{subsection}{0}}

\long\def\symbolfootnote[#1]#2{\begingroup 
\def\thefootnote{\fnsymbol{footnote}}\footnote[#1]{#2}\endgroup}

\def\be{\begin{equation}}      \def\ee{\end{equation}}
\def\bea{\begin{eqnarray}}     \def\eea{\end{eqnarray}}

\def\mb#1{\hbox{{\boldmath $#1$}}}

\highertp                   
\begin{document}
\sloppy

\begin{titlepage}
\addcontentsline{toc}{section}{Title Page}

\title{Four Lectures on Poincar\'e Gauge Field Theory\footnote{Given
at the 6th Course of the International School of Cosmology and
Gravitation on ``Spin, Torsion, Rotation, and Supergravity", held at
Erice, Italy, May 1979.}}
\author{Friedrich W. Hehl}
\date{}
\maketitle\thispagestyle{empty}

\begin{center}\vspace{-1.2cm}
Institute for Theoretical Physics,
University of Cologne, W. Germany\symbolfootnote[2]{Permanent
address.}\\
and\\
Center for Particle Theory and Center For Theoretical
Physics,\\ The University of Texas at Austin,\symbolfootnote[3]{Supported
in part by DOE contract DE-AS05-76ER-3992 and by NSF grant
PHY-7826592. }
Austin, Texas 78712
\end{center}

\centerline{\bf Abstract}\smallskip
{\small
The Poincar\'e (inhomogeneous Lorentz) group underlies special relativity.
In these lectures a consistent formalism is developed allowing an
appropriate gauging of the Poincar\'e group. The physical laws are formulated
in terms of points, orthonormal tetrad frames , and components of the matter
fields with respect to these frames. The laws are postulated to be gauge
invariant  under local Poincar\'e transformations. This implies the
existence of 4 translational gauge potentials $e^\alpha$ (``gravitons")
and 6 Lorentz gauge potentials ${\Gamma}^{\alpha\beta}$ (``rotons") and
the coupling of the momentum current and the spin current of matter to these
potentials, respectively. In this way one is led to a Riemann-Cartan spacetime
carrying torsion and curvature, richer in structure than the spacetime of general
relativity. The Riemann-Cartan spacetime is controlled by the two general gauge
field equations (\ref{344}) and (\ref{345}), in which material momentum and spin
act as sources. The general framework of the theory is summarized in a table in
Section 3.6. -- Options for picking a gauge field lagrangian are discussed
(telepa\-rallelism, ECSK). We propose a lagrangian quadratic in torsion and
curvature governing the propagation of gravitons and rotons. A suppression of
the rotons leads back to general relativity.}

\vfill\noindent{\small
{\bf Author's note.}
This work was originally published as a chapter in the book that is now long out
of print.\symbolfootnote[4]{\textcopyright\ Copyright (1980) Springer, reprinted
with kind permission from Springer Nature.\\
F. W. Hehl, Four Lectures on Poincar\'e Gauge Field Theory, in: P. G. Berg\-mann,
V. De Sabbata (eds), \emph{Cosmology and Gravitation}, NATO Advanced Study
Institutes Series, vol 58 (Springer, Boston, MA, 1980),
\url{https://doi.org/10.1007/978-1-4613-3123-0_2}.}
The purpose of this arXiv version is to make these lectures more accessible to
the current generation of students and researchers. I am extremely grateful to
my colleague and friend Milutin Blagojevi\'c (Belgrade) for arranging the
republication of my Erice lectures. Moreover, I'd like to thank his secretary
Vanja Mihajlovi\'c for putting the text most carefully into latex. For a more
modern look at that subject, see M.B. \& F.W.H. (eds.),
\href{https://doi.org/10.1142/p781}{Gauge Theories of Gravitation}, Imperial
College Press, London (2013).
}
\vspace{3pt}
\end{titlepage}

\newpage
\higher                          
\tableofcontents
\vfill
\setcounter{page}{2}
\lecture{Lecture 1: General Background}
\addcontentsline{toc}{section}{Lecture 1:~ General Background}

\subsection{Particle Physics and Gravity}

The recent development in particle physics seems to lead to the
following overall picture: the fundamental constituents of matter are
spin-on-half fermions, namely quarks and leptons, and their
interactions are mediated by gauge bosons coupled to the appropriate
conserved or partially conserved currents of the fermions. Strong,
electromagnetic, and weak interactions can be understood in this way
and the question arises, whether the gravitational interaction can be
formulated in a similar manner, too. These lectures are dedicated to
this problem.

General relativity is the most satisfactory gravitational theory so
far. It applies to macroscopic tangible matter and to electromagnetic
fields. The axiomatics of general relativity makes it clear that the
notions of massive test particles\symbolfootnote[1]{To be more precise:
massive, structureless, spherical symmetric, non-rotating, and neutral
test particles$\dots$.} and of massless scalar ``photons" underlie the
riemannian picture of spacetime. Accordingly, test particles, devoid of
any attribute other than mass-energy, trace the geodesics of the
supposed riemannian geometry of spacetime. This highly successful
conception of massive test particles and ``photons" originated from
classical particle mechanics and from the geometrical optics' limit of
electrodynamics, respectively. It is indispensable in the general
relativity theory of 1915 (GR).

Is it plausible to extrapolate riemannian geometry to microphysics? Or
shouldn't we rather base the spacetime geometry on the supposedly more
fundamental fermionic building blocks of matter?

\subsection{Local Validity of Special Relativity and Quantum Mechanics}

At least locally and in a suitable reference frame, special relativity
and quantum mechanics withstood all experimental tests up to the
highest energies available till now. Consequently we have to describe
an isolated particle according to the rules of special relativity and
quantum mechanics: Its state is associated with a unitary
representation of the Poincar\'e (inhomogeneous Lorentz) group. It is
characterized by its mass $m$ and by its  spin $s$. The universal
applicability of the mass-spin classification scheme to all known
particles establishes the Poincar\'e group as an unalterable element in
any approach to spacetime physics.

Let us assume then at first the doctrine of special relativity.
Spacetime is represented by a 4-dimensional differentiable manifold
$X_4$ the points of which are labelled by coordinates $x^i$. On the
$X_4$, a metric is given, and we require the vanishing of the
riemannian curvature. Then we arrive at a Minkowski spacetime $M_4$. We
introduce at each point an orthonormal frame of four vectors (tetrad
frame)
\be
\label{jedanjedan}
\til{e}_\alpha(x^k)=e^i_{\tck\alpha}{\partial_i}\quad
\textrm{with}\quad
\til{e}_\alpha\cdot \til{e}_\beta =\eta_{\alpha\beta}=
\textrm{diag.}\, (-+++)\ .
\ee
Here $\eta_{\alpha\beta}$ is the Minkowski metric.\symbolfootnote[1]{
The anholonomic (tetrad or Lorentz) indices $\alpha,\beta,\gamma \dots$
as well as the holonomic (coordinate or world) indices $i,j,k\dots$ run
from 0 to 3, respectively. For the notation and the conventions compare
\cite{1}. In the present article the object of anholonomity (\ref{15})
is defined with a factor 2, however. GR = general relativity of 1915,
PG = Poincar\'e gauge (field theory), P = Poincar\'e.} We have the dual
frame (co-frame) $\til{e}^\alpha=e_i^{\tck\alpha}dx^i$ and find
$e_i^{\tck\alpha} e^i_{\tck\beta}=\delta^\alpha_\beta$.

In the framework of the Poincar\'e gauge field theory (PG) to be
developed further down, the field of anholonomic tetrad frames
$\til{e}_\alpha(x^k)$ is to be considered an``irreducible" or primitive
concept. We imagine spacetime to be populated with observers. Each
observer is equipped with all the measuring apparatuses used in special
relativity, in particular with a length and an orientation standard
allowing him to measure spatial and temporal distances and relative
orientations, respectively. Such local observers are represented by the
tetrad field $\til{e}_\alpha(x^k)$. Clearly this notion of
``anholonomic observers" that lies at the foundations of the
PG,\symbolfootnote[2]{During the Erice school I distributed Kerlick's
translation of Cartan's original article \cite{2}. It should be clear
therefrom that it is Cartan who introduced this point of view.} is alien
to GR, as we saw above. It seems necessary, however, in order to
accommodate, at least at a local level, the experimentally well
established ``Poincar\'e behavior" of matter, in particular its
spinorial behavior.

\subsection{Matter and Gauge Fields}

After this general remark, let us come back to special relativity. In
the $M_4$ the global Poincar\'e group with its 10 infinitesimal
parameters (4 translations and 6 Lorentz-rotations) is the group of
motions. Matter, as mentioned, is associated with unitary
representations of the Poincar\'e group. The internal properties of
matter, the flavors and colors, will be neglected in our presentation
since we are only concerned with its spacetime behavior. Accordingly,
matter can be described by fields $\psi(x^k)$ which refer to the tetrad
$\til{e}_\alpha(x^k)$ and transform as Poincar\'e spinor-tensors,
respectively. Thereby, technically speaking, the $\psi(x^k)$ a priori
carry only anholonomic spinor and tensor indices, which we'll suppress
for convenience.

We will restrict ourselves to classical field theory, i.e. the fields
$\psi(x^k)$ are unquantized $c$-number fields. Quantization will have
to be postponed to later investigations.

The covariant derivative of a matter field reads
\be
D_i\psi(x^k)=(\partial_i+\Gamma^{\tck\alpha\beta}_i f_{\beta\alpha})
\psi(x^k)\ ,
\ee
where the $f_{\alpha\beta}$ are the appropriate constant matrices of
the Lorentz generators acting on $\psi(x^k)$. Their commutation
relations are given by
\be
\label{jedantri}
[f_{\alpha\beta}, f_{\gamma\delta}]=\eta_{\gamma[\alpha}
f_{\beta]\delta}-\eta_{\delta[\alpha}
f_{\beta]\gamma}\ .
\ee
The connection coefficients $\Gamma^{\tck\alpha\beta}_i$, being
referred to orthonormal tetrads on an $M_4$, can be expressed in terms
of the object of anholonomity $\Omega^{\tck\tck\alpha}_{ij}$
according to
\be
\label{jedan}
\Gamma_{\alpha\beta\gamma}: = e^i_{\tck\alpha}\eta_{\beta\delta}
\eta_{\gamma\varepsilon}\Gamma^{\tck\delta\varepsilon}_i=
-\frac{1}{2}\Omega_{\alpha\beta\gamma}+
\frac{1}{2}\Omega_{\beta\gamma\alpha}-
\frac{1}{2}\Omega_{\gamma\alpha\beta}
\ee
with
\be
\label{15}
\Omega_{\alpha\beta\gamma}: = e^i_{\tck\alpha}
e^j_{\tck\beta}
\eta_{\gamma\delta}
\Omega^{\tck\tck\delta}_{ij}\quad \textrm{and}\quad
\Omega^{\tck\tck\alpha}_{ij} : =2\partial_{[i}e^{\tck\alpha}
_{j]}\ .
\ee
We can read off from (\ref{jedan}) the antisymmetry of the connection
coefficients,
\be
\Gamma^{\tck\alpha\beta}_i \equiv -\Gamma^{\tck\beta\alpha}_i\ ,
\ee
i.e. neighboring tetrads are, apart from their relative displacement,
only rotated with respect to each other. Furthermore we
define and $\partial_\alpha : =e^i_{\tck\alpha}\partial_i$ and
$D_\alpha : = e^i_{\tck\alpha}D_i$.
For the mathematics involved we refer mainly to ref. \cite{3},
see also \cite{4}.

By definition, a field possessing originally a holonomic index, cannot
be a matter field. In particular, as it will turn out, gauge potentials
like the gravitational potentials $e_i^{\tck\alpha}$  and
$\Gamma^{\tck\alpha\beta}_i$ (see Section \ref{local}) or the
electromagnetic potential $A_i$, emerge with holonomic indices as
covariant vectors and do not represent matter
fields.\symbolfootnote[1]{Technically speaking gauge potentials are
always one-forms with values in some Lie-algebra, see O'Raifeartaigh
\cite{5}.} The division of physical fields into matter fields
$\psi(x^k)$ and gauge potentials like $e_i^{\tck\alpha}$,
$\Gamma^{\tck\alpha\beta}_i$, $A_i$ is natural and unavoidable in any
gauge approach (other than supergravity). In our gauge-theoretical
set-up, the gauge potentials and the associated fields will all be
presented by holonomic totally antisymmetric covariant tensors (forms)
or the corresponding antisymmetric contravariant tensor densities.
Hence there is no need of a covariant derivative for holonomic indices
and we require that the $D_i$ acts only on anholonomic indices, i.e.
\be
D_{[i}e_{j]}^{\,\tck\alpha}=\partial_{[i}e_{j]}^{\,\tck\alpha}+
\Gamma_{[i|\gamma}^{\,\tck\,\,\,\tck\alpha}
e_{|j]}^{\,\tck\gamma}\ ,
\ee
for example.\symbolfootnote[1]{Our $D_i$-operator (see \cite{1})
corresponds to the exterior covariant derivative of ref. \cite{4}}

We have seen that Poincar\'e matter is labelled by mass \emph{and}
spin. It is mainly this reason, why the description of matter by means
of a field should be superior to a particle description: The spin
behavior of matter can be better simulated in a field theoretical
picture. Additionally, already in GR, and in any gauge approach to
gravity, too, gravitation is represented by a field. Hence the
coherence of the theoretical model to be developed would equally
suggest a field-theoretical description of matter. After all, even in
GR matter dust is represented hydrodynamically, i.e.
field-theoretically. As a consequence, together with the notion of a
particle, the notion of a path, so central in GR, will loose its
fundamental meaning in a gauge approach to gravity. Operationally the
linear connection will then have to be seen in a totally different
context as compared to GR.\symbolfootnote[2] {In GR the holonomic
connection ${\tilde{\Gamma}}_{ij} ^{\tck\tck k}$ (the Christoffel) is
expressible in terms of the metric and has, accordingly, no independent
status. In the equation for the geodesics it represents a field
strength acting on test particles. In PG it is the \emph{an}holonomic
$\Gamma_i^{\tck\alpha\beta}$ which enters as a fundamental variable.
For its measurement we need a Dirac spin, see Section 3.3.} Only in a
macroscopic limit will we recover the conventional path concept again.

\subsection{Global Inertial Frames in the \texorpdfstring{\mb{M_4}}{M\_4} and
Action Function of Matter}

If we cover the $M_4$ with cartesian coordinates and orient all
tetrads parallely to the corresponding coordinate lines, then we find
trivially for the tetrad coefficients
\be
\label{18}
e_i^{\tck\alpha}\stackrel{\star}{=}\delta_i^\alpha\qquad
(e^j_{\tck\beta}\stackrel{\star}{=}\delta^j_\beta)\ ,
\ee
i.e., in the $M_4$ a we can build up global frames of reference,
inertial ones, of course. With respect to these frames,
the linear connection vanishes and we have for the corresponding
connection coefficients
\be
\label{19}
\Gamma_i^{\tck\alpha\beta}\stackrel{\star}{=} 0\ .
\ee
We will use these frames for the time being.

The lagrangian of the matter field will be assumed to be of
first order $L=L[\eta_{ij}, \gamma^i\dots,
\psi(x),\partial_i\psi(x)]$. The action function
reads
\be
\label{jdeset}
W_m=\int d^4xL(\eta_{ij},\gamma^i\dots,\psi,\partial_i\psi)\ ,
\ee
where $\gamma^i$ denotes the Dirac matrices, e.g.
The invariance of (\ref{jdeset}) under global Poincar\'e
transformations yields momentum and angular
momentum conservation, i.e., we find a conserved momentum current
(energy-momentum tensor) and a conserved angular momentum current.

\subsection{Gauging the Poincar\'e Group and Gravity}

Now the gauge idea sets in. Global or rigid Poincar\'e invariance is of
questionable value. From a field-theoretical point of view, as first
pointed out by Weyl \cite{6} and Yang and Mills \cite{7}, and applied
to gravity by Utiyama \cite{8}, Sciama \cite{9} and Kibble \cite{10},
it is unreasonable to execute at each point of spacetime the same rigid
transformation. Moreover, what we know experimentally, is the existence
of minkowskian metrics all over. How these metrics are oriented with
respect to each other, is far less well known, or, in other words,
\emph{local} Poincar\'e invariance is really what is observed.
Spacetime is composed of minkowskian granules, and we have to find out
their relative displacements and orientations with respect to each
other.

Consequently we substitute the $(4+6)$ infinitesimal parameters of a
Poincar\'e transformation by $(4+6)$ spacetime-dependent functions and
see what we can do in order to save the invariance of the action
function under these extended, so-called \emph{local} Poincar\'e
transformations. (We have to introduce $(4+6)$ compensating vectorial
gauge potentials, see Lecture 2.)

This brings us back to gravity. According to the equivalence principle,
there exists in GR in a freely falling coordinate frame the concept of
the local validity of special relativity, too. Hence we see right away
that gauging the Poincar\'e group must be related to gravitational
theory. This is also evident from the fact that, by introducing local
Poincar\'e invariance, the conservation of the momentum current is at
disposition, inter alia. Nevertheless, the gauge-theoretical
exploitation of the idea of a local Minkowski structure leads to a more
general spacetime geometry, namely to a Riemann-Cartan or $U_4$
geometry, which seems to be at conflict with Einstein's result of a
riemannian geometry. The difference arises because Einstein, in the
course of heuristically deriving GR, treats material particles as
described in holonomic coordinate systems, whereas we treat matter
fields which are referred to anholonomic tetrads.

These lectures cover the basic features of the Poincar\'e gauge field
theory (``Poincar\'e gauge", abbreviated PG). Our outlook is strictly
\emph{phenomenological}, hopefully in the best sense of the word. For a
list of earlier work we refer to the review article \cite{1}. The
articles of Ne'eman \cite{11}, Trautman \cite{12} and Hehl, Nitsch, and
von der Heyde \cite{13} in the Einstein Commemorative Volume together
with information from the lectures and seminars given here in Erice by
Ne'eman \cite{14}, Trautman \cite{15}, Nitsch \cite{16}, Rumpf
\cite{17}, W. Szczyrba \cite{18}, Schweizer \cite{19}, Yasskin
\cite{20}, and by ourselves, should give a fairly complete coverage of
the subject. But one should also consult Tunyak \cite{21}, who wrote a
whole series of most interesting articles, the Festschrift for Ivanenko
\cite{22} where earlier references of Ivanenko and associates can be
traced back, and Zuo et al \cite{23}.

\lecture{Lecture 2: Geometry of Spacetime}
\addcontentsline{toc}{section}{Lecture 2:~ Geometry of Spacetime}

We have now an option. We can either start with an $M_4$ and substitute
the parameters in the P(oincar\'e)-transformation of the matter fields
by spacetime dependent functions and work out how to compensate the
invariance violating terms in the action function: this was carried
through in ref. \cite{1}, where it was shown in detail how one arrives
at a $U_4$ geometry with torsion and curvature. Or, following von der
Heyde \cite{24}, \cite{25}, we can alternatively postulate a
\emph{local} P-structure everywhere on an $X_4$, derive therefrom in
particular the transformation properties of the gauge potentials, and
can subsequently recover the global P-invariance in the context of an
$M_4$ as a special case. Both procedures lead to the same results. We
shall follow here the latter one.

\subsection{Orthonormal Tetrad Frames and Metric
Com\-pa\-tible Connection}

On an $X_4$ let there be given a sufficiently differentiable  field of
tetrad frames $\til{e}_\alpha(x^k)$. Additionally, we assume the
existence of a Minkowski metric $\eta_{\alpha\beta}$. Consequently,
line in (\ref{jedanjedan}), we can choose the tetrad to be orthonormal,
furthermore we can determine $e^i_{\tck\alpha}$, $e_i^{\tck\alpha}$,
and $e:=\textrm{det}\, e_i^{\tck\alpha}$. The relative \emph{position}
of an event with respect to the origin of a tetrad frame is given by
$dx^\alpha=e_i^{\tck\alpha}dx^i$ and the corresponding distance by
$ds=(\eta_{\alpha\beta} dx^\alpha dx^\beta)^{1/2}$.

Let also be given a local standard of \emph{orientation}. Then,
starting from a tetrad frame $\til{e}_\alpha(x^k)$, we are able to
construct, at a point infinitesimally nearby, a parallelly oriented
tetrad
\be
\label{dvajedan}
\til{e}_\alpha^{''}(x^k+dx^k)=\til{e}_\alpha(x^k)+dx^i\Gamma
_{i\alpha}^{\tck\tck\beta}(x^k)\til{e}_\beta(x^k)\ ,
\ee
provided the connection coefficients $\Gamma
_{i\alpha}^{\tck\tck\beta}$ are given.

The metric, and thereby the length standard, are demanded to be defined
globally, i.e. lengths and angles must stay the same under parallel
transport:
\be
\label{dvadva}
\til{e}_\alpha^{''}(x^k+dx^k)\cdot\til{e}_\beta^{''}(x^k+dx^k)=
\til{e}_\alpha(x_k)\cdot \til{e}_\beta(x^k)=\eta_{\alpha\beta}\,.
\ee
Upon substitution of (\ref{dvajedan}) into (\ref{dvadva}), we find a
metric compatible connection\symbolfootnote[1]{Observe that
$\Gamma_i^{\tck\alpha\beta}$ now represents an independent variable, it
is no longer of the type as given in eq. (\ref{jedan}). P-gauge
invariance requires the existence of an independent rotational
potential, see ref. \cite{1}.}
\be
\label{23}
\Gamma_i^{\tck\alpha\beta}=-\Gamma_i^{\tck\beta\alpha}\ .
\ee
The $(16+24)$ independent quantities $(e_i^{\tck\alpha},
\Gamma_i^{\tck\alpha\beta})$ will be the variables of our
theory.\symbolfootnote[2]{Instead of $e_i^{\tck\alpha}$, we could also
use $e^j_{\tck\beta}$ as independent variable. This would complicate
computations, however. We know from electrodynamics that the gauge
potential is a covariant vector (one-form) as is the rotational
potential $\Gamma_i^{\tck\alpha\beta}$. Then $e_i^{\tck\alpha}$, as a
covariant vector, is expected to be more suitable as a gauge potential
than $e^j_{\tck\beta}$, and exactly this shows up in explicit
calculations.} The anholonomic metric $\eta_{\alpha\beta}$ is a
constant, the holonomic metric
\be
g_{ij}: = e_i^{\tck\alpha}e_j^{\tck\beta}\eta_{\alpha\beta}
\ee
a convenient abbreviation with no independent status.

The total arrangement of all tetrads with respect to each other in
terms of their relative positions and relative orientations makes up
the geometry of spacetime. Locally we can only recognize a special
relativistic structure. If the global arrangement of the tetrads with
respect to position and orientation is integrable, i.e.
path-independent, then we have an $M_4$, otherwise a non-minkowskian
spacetime, namely a $U_4$ or Riemann-Cartan spacetime.

\subsection{Local P-Transformation of the Matter Field}

We base our considerations on an active interpretation of the
P-transfor\-mation. We imagine that the tetrad field and the coordinate
system are kept fixed, whereas the matter field is ``transported". A
matter field $\psi(x^k)$, being translated from $x^k$ to $x^k +
\varepsilon^k$, where $\varepsilon^k$ are the 4 infinitesimal
parameters of translations and
$\varepsilon^\gamma=e_k^{\tck\gamma}\varepsilon^k$, has to keep its
orientation fixed and, accordingly, the generator of translations is
that of a parallel transport,\symbolfootnote[3]{For this reason,
Ne'man's title of this article \cite{26} reads: ``Gravity is the gauge
theory of the parallel-transport modification of the Poincar\'e
group".} i.e. it is the covariant derivative operator
\be
\label{25}
D_\gamma\psi(x)=e^i_{\tck\gamma}D_i\psi(x)=e^i_{\tck\gamma}
(\partial_i+\Gamma_i^{\tck\alpha\beta}f_{\beta\alpha})\psi(x)\ .
\ee
It acts only on anholonomic indices, see the analogous discussion
in Section 1.3. This transformation of a translational type
distinguishes the PG from gauge theories for internal symmetries,
since the matter field is shifted to a different point in spacetime.

\begin{figure}[htb]
\centering
\includegraphics[height=5cm]{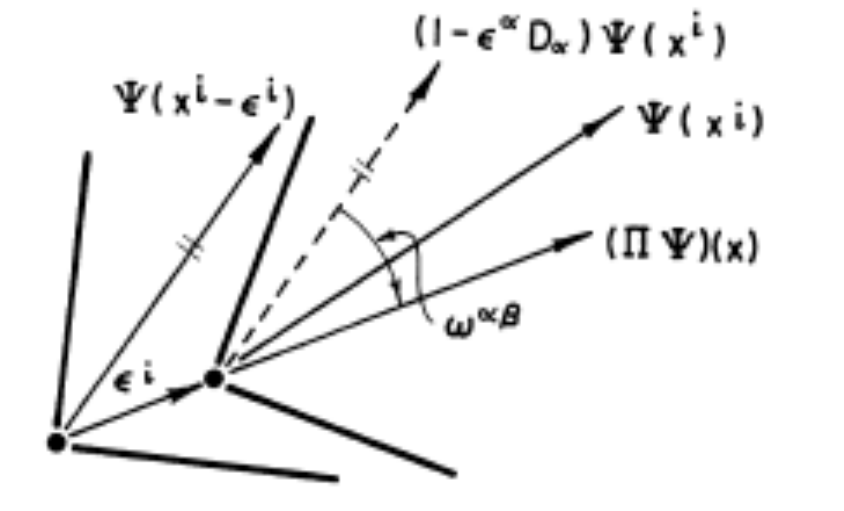}
\vspace{-.3cm}
\caption{An infinitesimal active local Poincar\'e
transformation of a matter field: The field $\psi(x^i-\varepsilon^i)$
is first parallelly displaced over the infinitesimal vector
$\varepsilon^i=e^i_{\tck\gamma}\varepsilon^\gamma$, rotated by the
angle $\omega^{\alpha\beta}$, and then compared with $\psi(x^i)$.}
\end{figure}

The Lorentz-rotations (6 infinitesimal parameters
$\omega^{\alpha\beta}$) are of the standard, special relativistic type.
Hence the local P-transformation $\Pi$ of a field reads (see Figure 1):
\be
\label{26}
(\Pi\psi)(x)=(1-\varepsilon^\gamma(x)D_\gamma+\omega^{\alpha\beta}
(x)f_{\beta\alpha})\psi(x)\ .
\ee
Here again, the $f_{\beta\alpha}$ are the matrices of the Lorentz
generators obeying (\ref{jedantri}). Of course, setting up a gauge
theory, the $(4+6)$ infinitesimal ``parameters" $(\varepsilon^\gamma,
\omega^{\alpha\beta})$ are spacetime dependent functions. A matter
field distribution $\psi(x)$, such is our postulate, after the
application of a local P-transformation $\Pi$, i.e.
$\psi(x)\to(\Pi\psi)(x)$, is equivalent in all its measurable
properties to the original distribution $\psi(x)$.

\subsection{Commutation Relations, Torsion, and Curvature}

The translation generators $D_\gamma$ and the rotation generators
$f_{\beta\alpha}$ fulfill commutation relations which we will derive
now. The commutation relations for the $f_{\beta\alpha}$ with
themselves are given by the special relativistic formula
(\ref{jedantri}). For rotations and translations we start with the
relation $f_{\alpha\beta}D_i=D_if_{\alpha\beta}$, which is valid since
$D_i$ doesn't carry a tetrad index. Let us remind ourselves that the
$f_{\alpha\beta}$, as operators, act on everything to their rights.
Transvecting with $e^i_{\tck\gamma}$, we find
$e^i_{\tck\gamma}f_{\alpha\beta}D_i=D_\gamma f_{\alpha\beta}$, i.e.,
\be
\label{27}
[f_{\alpha\beta},D_\gamma]=[f_{\alpha\beta},e^i_{\tck\gamma}]
D_i=\eta_{\gamma[\alpha}D_{\beta]}\ .
\ee
This formula is strictly analogous to its special relativistic
pendant.

Finally, let us consider the translations under themselves.
By explicit application of (\ref{25}), we find
\be
\label{28}
[D_i,D_j]=F_{ij}^{\tck\tck\alpha\beta}f_{\beta\alpha}\ ,
\ee
where
\be
\label{29}
F_{ij\alpha}^{\tck\tck\tck\beta}: =
2\bigg(
\partial_{[i}\Gamma_{j]\alpha}^{\tck \ \tck\beta}
+\Gamma_{[i|\gamma}^{\tck \ \tck\beta}\Gamma_{|j]\alpha}
^{\,\tck \ \tck\gamma}
\bigg)
\ee
is the curvature tensor (rotation field strength). Now
$D_i=e_i^{\tck\alpha}D_\alpha$, substitute it into (\ref{28}) and
define the torsion tensor (translation field strength),
\be
\label{210}
F_{ij}^{\tck\tck\alpha}: =2D_{[i}e_{j]}^{\tck\alpha}=
2\bigg(\partial_{[i}e_{j]}^{\tck\alpha}
+\Gamma_{[i|\beta}^{\tck \ \tck\alpha}e_{|j]}^{\,\tck\beta}
\bigg)
\ee
Then finally, collecting all relevant commutation relations, we have
\bea
\label{211}
&&[D_\alpha,D_\beta] = -F_{\alpha\beta}^{\tck\tck\gamma}
  D_\gamma+F_{\alpha\beta}^{\tck\tck\gamma\delta}
  f_{\delta\gamma}\ ,                                   \\
\label{212}
&&[f_{\alpha\beta},D_\gamma] = \eta_{\gamma[\alpha}D_{\beta]}\ ,\\
\label{213}
&&[f_{\alpha\beta},f_{\gamma\delta}] =
  \eta_{\gamma[\alpha}f_{\beta]\delta}-
  \eta_{\delta[\alpha}f_{\beta]\gamma}\ .
\eea

For vanishing torsion and curvature we recover the commutation
relations of global P-transformations. \emph{Local} P-transformations,
in contrast to the corresponding structures in gauge theories of
internal symmetries, obey different commutation relations, in
particular the algebra of the translations doesn't close in general.
Observe, however, that it does close for vanishing curvature, i.e. in a
spacetime with teleparallelism (see Section 2.7). In introducing our
translation generators, we already stressed their unique features. This
is now manifest in (\ref{211}). Note also that the ``mixing term"
$2\Gamma_{[i|\beta}^{\tck \ \tck\alpha}e_{|j]} ^{\,\tck\beta}$ between
translational \emph{and} rotational potentials in the'definition
(\ref{210}) of the translation field is due to the existence of orbital
angular momentum.

In deriving (\ref{211}), we find in torsion and curvature the tensors
which covariantly characterize the possibly different arrangement of
tetrads in comparison with that in special relativity. Torsion and
curvature measure the non-minkowskian behavior of the tetrad
arrangement. In (\ref{211}) they relate to the translation and rotation
generators, respectively. Consequently torsion represents the
translation field strength and curvature the rotation field strength.

The only non-trivial Jacobi identity,
\be
\label{214}
[D_\alpha,[D_\beta,D_\gamma]]+[D_\beta,[D_\gamma,D_\alpha]]
+[D_\gamma,[D_\alpha,D_\beta]]=0\ ,
\ee
leads, after substitution of (\ref{211}), some algebra and using
(\ref{244}), to the two sets of Bianchi identities
\bea
\label{215}
&&D_{[i}F_{jk]}^{\tck\tck\,\beta}\equiv
  F_{[ijk]}^{\tck\tck\tck\,\beta}\ ,                 \\
\label{216}
&&D_{[i}F_{jk]\alpha}^{\tck\tck\,\tck\beta}\equiv 0\ .
\eea

\subsection{Local P-Transformation of the Gauge Potentials}
\label{local}

Let us now come back to our postulate of local P-invariance. A matter
field distribution actively P-transformed, $\psi(x)\to(\Pi\psi)(x)$ ,
should be equivalent to $\psi(x)$ . How can it happen that a local
observer doesn't see a difference in the field configuration after
applying the P-transformation? The local P-transformation will induce
not only a variation of $\psi(x)$ , but also correct the values of the
tetrad coefficients $e_i^{\tck\alpha}$ and the connection coefficients
$\Gamma_i^{\tck\alpha\beta}$ such that a difference doesn't show up. In
other words, the local P-transformation adjusts suitably the relative
position and the relative orientation of the tetrads as determined by
the corresponding coefficients ($e_i^{\tck\alpha},
\Gamma_i^{\tck\alpha\beta}$). Thereby the P-transformation of the gauge
potentials is a consequence of the local P-structure of spacetime.

\begin{figure}[htb]
\centering
\includegraphics[height=6cm]{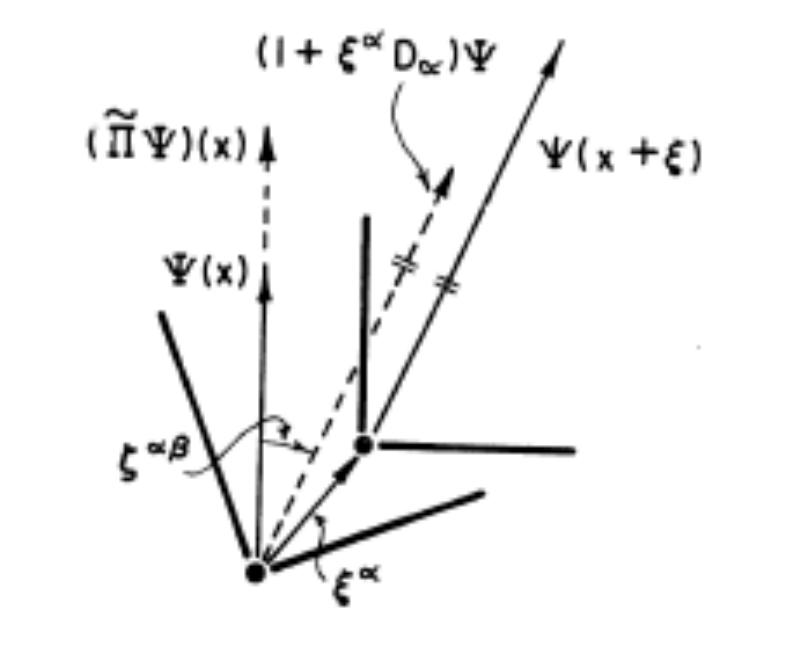}\vspace{-.5cm}
\caption{A matter field distribution near $x^k$.}
\end{figure}

Consider a matter field distribution $\psi(x)$, in particular its
values at $x^k$ and at a nearby point $x^k + \xi^k$. See Figure 2 where
the matter field is symbolized by a vector. The relative position of
$\psi(x+\xi)$ and $\psi(x)$ is determined by $\xi^\alpha$, their
relative orientation by $\zeta^{\alpha\beta}=\zeta^{[\alpha\beta]}$,
the angle between $(1 + \xi^\alpha D_\alpha)\psi(x)$ and $\psi(x)$. By
a rotation $-\zeta^{\alpha\beta}$ of $(1+\xi^\alpha D_\alpha)\psi(x)$,
we get $(\tilde{\Pi}\psi)(x)=(1+\xi^\alpha D_\alpha-
\zeta^{\alpha\beta}f_{\beta\alpha})\times
{\mathop{\psi}}(x)$, which is, of course, parallel to
$\psi(x)$. The transformation $\tilde{\Pi}$ has the same structure as a
P-transformation.\symbolfootnote[1] {It is not a P-transformation,
since we consider the untransformed matter field distribution.}

Now P-transform $\psi(x)$ and $\psi(x + \xi)$. Then $(\tilde{\Pi}\psi)
(x)$ must stay parallel to $\psi(x)$, i.e. it is required to transform
as a P-spinor-tensor. Furthermore $(\Pi\psi)(x+\xi)$ and
$(\Pi\psi)(x)$, the P-transforms of the fields $\psi(x+\xi)$ and
$\psi(x)$, can be again related by a transformation of the type
$\tilde\Pi$, i.e. there emerge new $\xi^\alpha$ and
$\zeta^{\alpha\beta}$ which are the P-transforms of the old ones.

Consequently, $D_\alpha\psi$ as well as $f_{\alpha\beta}\psi$
transform as P-spinor-tensors, i.e. if, according to
(\ref{26}) $\psi\to\Pi\psi$, then
\bea
\label{217}
&&D_\alpha\psi\to\Pi D_\alpha\psi\ ,            \\
\label{218}
&&f_{\alpha\beta}\psi\to\Pi f_{\alpha\beta}\psi\ .
\eea
This implies, via the commutation relations (\ref{211}), that torsion
$F_{\alpha\beta}^{\tck\tck\gamma}$ and curvature
$F_{\alpha\beta}^{\tck\tck\gamma\delta}$ are also P-tensors, an
information which will turn out to be useful in constructing invariant
gauge field Lagrangians.

Now we have
\be
\label{219}
\Pi D_\alpha\psi=(D_\alpha+\delta D_\alpha)(\psi+\delta\psi)=
(D_\alpha+\delta D_\alpha)\Pi\psi
\ee
and, because $\Pi$ deviates only infinitesimally from unity,
\be
\label{220}
\delta D_\alpha=[\Pi,D_\alpha]\ .
\ee
Analogously we find
\be
\label{221}
\delta f_{\alpha\beta}=[\Pi,f_{\alpha\beta}]\ .
\ee
Substitution of $\Pi$ from (\ref{26}) into (\ref{220}), (\ref{221})
and using the commutation relations (\ref{211})-(\ref{213}), yields,
respectively,
\bea
\label{222}
\delta D_\alpha &=& -[\varepsilon^\gamma D_\gamma,D_\alpha]+
[\omega^{\gamma\delta}f_{\delta\gamma},D_\alpha]\nonumber\\
&=&-\Big(-D_\alpha\varepsilon^\gamma+\omega_\alpha^{\tck\gamma}
-\varepsilon^\delta F_{\delta\alpha}^{\tck\tck\gamma}\Big)
D_\gamma\nonumber\\
&&+
\Big(-D_\alpha\omega^{\tck\gamma}_\varepsilon -
\varepsilon^\delta F_{\delta\alpha\varepsilon}
^{\tck\tck\tck\gamma}\Big)f_\gamma^{\tck\varepsilon}\ ,\\
\label{223}
\delta f_{\alpha\beta}&=&-[\varepsilon^\gamma D_\gamma,f_{\alpha\beta}]
+[\omega^{\gamma\delta}f_{\delta\gamma},f_{\alpha\beta}]=0\ .
\eea
The coordinates are kept fixed during the active P-transformation. Then,
using (\ref{25}) and (\ref{223}), we get
\be
\label{224}
\delta D_i=\delta\Big(\partial_i+\Gamma_i^{\tck\alpha\beta}
f_{\beta\alpha}\Big)=\Big(\delta\Gamma_i^{\tck\alpha\beta}\Big)
f_{\beta\alpha}
\ee
or
\be
\label{225}
\delta D_\alpha=\delta\Big(e^i_{\tck\alpha}D_i\Big)=
\Big(\delta e^i_{\tck\alpha}\Big)D_i+
e^i_{\tck\alpha}\Big(
\delta\Gamma_i^{\tck\alpha\beta}\Big)
f_{\beta\alpha}\ .
\ee
A comparison with (\ref{222}) and remembering
$\delta(e^i_{\tck\alpha}e_j^{\tck\alpha})=0$, yields the desired
relations\symbolfootnote[1] {It might be interesting to note that, in 3
dimensions, these relations represent essentially the two
\emph{deformation} tensors of a so-called Cosserat continuum as
expressed in terms of their translation fields $\varepsilon^\alpha$ and
rotation fields $\omega^{\alpha\beta}$. Those analogies suggested to us
at first the existence of formulas of the type (\ref{226}),
(\ref{227}). They were first proposed in ref. \cite{1}, cf. also refs.
\cite{27}, \cite{28}.}
\bea
\label{226}
&&\delta e_i^{\tck\alpha}=-D_i\varepsilon^\alpha
  +\omega_\gamma^{\tck\alpha}e_i^{\tck\gamma}
  -\varepsilon^\gamma F_{\gamma i}^{\tck\tck\alpha}\ ,     \\
\label{227}
&&\delta\Gamma_i^{\tck\alpha\beta}=-D_i\omega^{\alpha\beta}
  -\varepsilon^\gamma F_{\gamma i}^{\tck\tck\alpha\beta}\,.
\eea

From gauge theories on internal groups we are just not used
to the non-local terms carrying the gauge field strengths; this is
again an outflow of the specific behavior of the translations.
Otherwise, the first terms on the right hand sides of (\ref{226}),
(\ref{227}), namely $-D_i\varepsilon^\alpha$ and
$-D_i\omega^{\alpha\beta}$, are standard. They express the
nonhomogeneous transformation behavior of the potentials under local
P-gauge transformations, respectively, and it is because of this fact
that the names translation and rotation gauge potential are justified.
The term $(\omega_\gamma^{\tck\alpha}e_i^{\tck\gamma})$ in (\ref{226})
shows that the tetrad, the translation potential, behaves as a
\emph{vector} under rotations. This leads us to expect, as indeed will
turn out to be true, that $e_i^{\tck\alpha}$, in contrast to the
rotation potential $\Gamma_i^{\tck\alpha\beta}$, should carry intrinsic
spin.

Having a $U_4$ with torsion and curvature we know that besides local
P-invariance, we have additionally invariance under general
\emph{coordinate} transformations. In fact, this coordinate invariance
is also a consequence of our formalism (see \cite{1}). Starting with a
$U_4$, one can alternatively develop a ``gauge" formalism with
coordinate invariance and local Lorentz invariance as applied to
tetrads as ingredients. The only difference is, however, that with our
procedure we recover in the limiting case of an $M_4$ exactly the
well-known global P-transformations of special relativity (see eq.
(\ref{260})) including the corresponding conservation laws (see eqs.
(\ref{312}), (\ref{313})), whereas otherwise the global gauge limit in
this sense is lost: We have only to remember that in special relativity
energy-momentum conservation is \emph{not} a consequence of coordinate
invariance, but rather of invariance under
translations.\symbolfootnote[1] {Such an alternative formalism was
presented by Dr. Schweizer in his highly interesting seminar talk
\cite{19}. His tetrad loses its position as a potential, since it
transforms homogeneously under coordinate transformations. Furthermore,
having got rid of the $M_4$-limit as discussed above, one has to be
very careful about what to define as local Lorentz-invariance.
Schweizer defines \emph{strong} as well as \emph{weak} local
Lorentz-invariance. However, the former notion lacks geometrical
significance altogether. Whereas we agree with Schweizer that there is
nothing mysterious about the local P-gauge approach and that one can
readily rewrite it in terms of coordinate and Lorentz-invariance (cf.
\cite{1}, Sect. IV. C. 3) and with the help of Lie-derivatives (cf.
\cite{29}), we hold that in our formulation there is a completely
satisfactory place for the translation gauge (see also
\cite{30,26,31,32}). Hence we leave it to others ``to gauge the
translation group more attractively..."}

\subsection{Closure of the Local P-Transformations}

Having discussed so far how the matter field $\psi(x)$ and the gauge
potentials $e^{\tck\alpha}_i$ and $\Gamma_i^{\tck\alpha\beta}$
transform under local P-transformations, we would now like to show once
more the intrinsic naturality and usefulness of the local P-formalism
developed so far. We shall compute the commutator of two successive
P-transformations $[\stackrel{2}{\Pi},\stackrel{1}{\Pi}]$. We will find
out that it yields a third local P-transformation $\stackrel{3}{\Pi}$,
the infinitesimal parameters ($\stackrel{3}{\varepsilon}{}^\alpha,
\stackrel{3}{\omega}{}^{\alpha\beta}$) of which depend in a suitable
way on the parameters ($\stackrel{1}{\varepsilon}{}^\alpha,
\stackrel{1}{\omega}{}^{\alpha\beta}$) and ($\stackrel{2}{\varepsilon}
{}^\alpha, \stackrel{2}{\omega}{}^{\alpha\beta}$) of the two
transformations $\stackrel{1}{\Pi}$ and $\stackrel{2}{\Pi}$,
respectively.\symbolfootnote[1]{The proof was first given by Nester
\cite{29}. Ne'eman and Takasugi \cite{33} generalized it to
supergravity including the ghost regime.}

Take the connection as an example. We have
\be
\label{228}
\stackrel{1}{\Gamma}{}^{\tck\alpha\beta}_i: =
\stackrel{1}{\Pi}\Gamma_i^{\tck\alpha\beta}=
\Gamma_i^{\tck\alpha\beta}-D_i\stackrel{1}{\omega}
{}^{\alpha\beta}-\stackrel{1}{\varepsilon}{}^\gamma
e_i^{\tck\delta}F_{\gamma\delta}^{\tck\tck\alpha\beta}\ .
\ee
In the last term we have purposely written the curvature in its totally
anholonomic form. Then it is a P-tensor, as we saw in the last section.
Applying now $\stackrel{2}{\Pi}$, we have to keep in mind that
$\stackrel{2}{\Pi}$ acts with respect to the transformed tetrad
coefficients $\stackrel{1}{e}{}^{\tck\delta}_i: =
\stackrel{1}{\Pi}e_i^{\tck\delta}$ as well as with respect to the
transformed connection coefficients
$\stackrel{1}{\Gamma}{}_i^{\tck\alpha\beta}$. Consequently we have
\be
\label{229}
\stackrel{2}{\Pi}\stackrel{1}{\Gamma}{}^{\tck\alpha\beta}_i=
\stackrel{2}{\Pi}\stackrel{1}{\Pi}\Gamma{}^{\tck\alpha\beta}_i=
\stackrel{1}{\Gamma}{}^{\tck\alpha\beta}_i-\stackrel{1}{D}_i
\stackrel{2}{\omega}{}^{\alpha\beta}-
\stackrel{2}{\varepsilon}{}^\gamma
\stackrel{1}{e}{}^{\tck\delta}_i
\stackrel{1}{F}{}_{\gamma\delta}^{\tck\tck\alpha\beta}\ .
\ee

Now we will evaluate the different terms in (\ref{229}). By
differentiation and by use of (\ref{228}) we find
\bea
\label{230}
\stackrel{1}{D}_i
\stackrel{2}{\omega}{}^{\alpha\beta}&=&
D_i\stackrel{2}{\omega}{}^{\alpha\beta}+2
\Big(
\stackrel{1}{\delta}\Gamma_{i\mu}^{\tck\tck[\alpha|}\Big)
\stackrel{2}{\omega}{}^{\mu|\beta]}\nonumber\\
&=&
{D}_i
\stackrel{2}{\omega}{}^{\alpha\beta}
-2
\Big(D_i\stackrel{1}{\omega}{}_\mu^{\tck[\alpha|}\Big)
\stackrel{2}{\omega}{}^{\mu|\beta]}\nonumber\\
&&-2\stackrel{1}{\varepsilon}{}^\gamma
F_{\gamma i\mu}^{\tck\tck\tck[\alpha|}
\stackrel{2}{\omega}{}^{\mu|\beta]}\,.
\eea

Let us turn to the last term in (\ref{229}). If we apply
(\ref{226}), it reads in the appropriate order:

\bea
\label{231}
\stackrel{2}{\varepsilon}{}^\gamma\stackrel{1}{e}{}^{\tck\delta}_i
\stackrel{1}{F}{}_{\gamma\delta}^{\tck\tck\alpha\beta}
&=&
\stackrel{2}{\varepsilon}{}^\gamma e^{\tck\delta}_i
\stackrel{1}{F}{}_{\gamma\delta}^{\tck\tck\alpha\beta}+
\stackrel{2}{\varepsilon}{}^\gamma\Big(
\stackrel{1}{\delta}e^{\tck\delta}_i\Big)
{F}_{\gamma\delta}^{\tck\tck\alpha\beta}\nonumber\\
&=& \stackrel{2}{\varepsilon}{}^\gamma
e^{\tck\delta}_i
\stackrel{1}{F}{}_{\gamma\delta}^{\tck\tck\alpha\beta}
+
\stackrel{2}{\varepsilon}{}^\gamma\nonumber\\
&&\Big(
-D_i\stackrel{1}{\varepsilon}{}^\delta+
\stackrel{1}{\omega}{}^{\tck\delta}_\nu
e^{\tck\nu}_i-\stackrel{1}{\varepsilon}{}^\nu
{F}_{\nu i}^{\tck\tck\delta}\Big)
{F}_{\gamma\delta}^{\tck\tck\alpha\beta}\ .
\eea
In (\ref{231}) there occurs the P-transform of the anholonomic
curvature. Like any P-tensor, it transforms according to (\ref{26}):
\be
\label{232}
\stackrel{1}{F}{}_{\gamma\delta}^{\tck\tck\alpha\beta}=
F_{\gamma\delta}^{\tck\tck\alpha\beta}-
\stackrel{1}{\varepsilon}{}^\mu D_\mu
F_{\gamma\delta}^{\tck\tck\alpha\beta}-
2\stackrel{1}{\omega}{}_{[\gamma|}^{\ \tck \ \mu}
F_{\mu|\delta]}^{\tck\,\tck\, \alpha\beta}+
2\stackrel{1}{\omega}{}_\mu^{\tck[\alpha|}
F_{\gamma\delta}^{\tck\tck\mu|\beta]}\ .
\ee

Now we substitute first (\ref{232}) into (\ref{231}). The
resulting equation together with (\ref{228}) and (\ref{230}) are then
substituted into (\ref{229}). After some reordering we find

\bea
\label{233}
\stackrel{2}{\Pi}\stackrel{1}{\Pi}\Gamma_i^{\tck\alpha\beta}&=&
\Gamma_i^{\tck\alpha\beta}-D_i\Big(
\stackrel{1}{\omega}{}^{\alpha\beta}+
\stackrel{2}{\omega}{}^{\alpha\beta}\Big)-
F_{\gamma i}^{\tck\tck\alpha\beta}\Big(
\stackrel{1}{\varepsilon}{}^\gamma+\stackrel{2}{\varepsilon}{}^\gamma
\Big)+\nonumber\\
&&+
F_{\gamma\delta}^{\tck\tck\alpha\beta}
F_{\nu i}^{\tck\tck\delta}\stackrel{1}{\varepsilon}{}^\nu
\stackrel{2}{\varepsilon}{}^\gamma+2D_i\Big(
\stackrel{1}{\omega}{}_\mu^{\tck[\alpha|}\Big)
\stackrel{2}{\omega}{}^{\mu|\beta]}+\nonumber\\
&&+
2F_{\gamma i\mu}^{\tck\tck\tck[\alpha|}\Big(
\stackrel{1}{\varepsilon}{}^\gamma\stackrel{2}{\omega}{}^{\mu|\beta]}
+\stackrel{2}{\varepsilon}{}^\gamma
\stackrel{1}{\omega}{}^{\mu|\beta]}\Big)+e^{\tck\delta}_i
F_{\mu\delta}^{\tck\tck\alpha\beta}
\stackrel{2}{\varepsilon}{}^\gamma
\omega_\gamma^{\tck\mu}+\nonumber\\
&&+
e^{\tck\delta}_i\Big(D_\mu
F_{\gamma\delta}^{\tck\tck\alpha\beta}\Big)
\stackrel{2}{\varepsilon}{}^\gamma
\stackrel{1}{\varepsilon}{}^\mu +
F_{\gamma\delta}^{\tck\tck\alpha\beta}
\stackrel{2}{\varepsilon}{}^\gamma D_i
\stackrel{1}{\varepsilon}{}^\delta\ .
\eea

I am not happy myself with all those indices. Anybody is invited to
look for a simpler proof. But the main thing is done. The
right-hand-side of (\ref{233}) depends only on the untransformed
geometrical quantities and on the parameters. By exchanging the one's
and two's wherever they appear, we get the reversed order of the
transformations:
\be
\label{234}
\stackrel{1}{\Pi}\stackrel{2}{\Pi}\Gamma_i^{\tck\alpha\beta}=
(\ref{233})\quad\textrm{with}\quad
\begin{array}{rl}
1\to2\\
2\to1
\end{array}\ .
\ee
Subtracting (\ref{234}) from (\ref{233}) yields the commutator
$[\stackrel{2}{\Pi}\stackrel{1}{\Pi}]\Gamma_i^{\tck\alpha\beta}$.

After some heavy algebra and application of the 2nd Bianchi identity
in its anholonomic form,
\be
\label{235}
D_{[\alpha}F_{\beta\gamma]}^{\tck\tck\mu\nu}=
F_{[\alpha\beta}^{\tck\,\tck\delta}F_{\gamma]\delta}
^{\tck \, \tck\mu\nu}\ ,
\ee
we find indeed a transformation $\stackrel{3}{\Pi}
\Gamma_i^{\tck\alpha\beta}$ of the form (\ref{228}) with the
following parameters:
\bea
\label{236}
&&\stackrel{3}{\varepsilon}{}^\alpha=-
  \stackrel{2}{\omega}{}^{\tck\alpha}_\gamma
  \stackrel{1}{\varepsilon}{}^\gamma+
  \stackrel{1}{\omega}{}^{\tck\alpha}_\gamma
  \stackrel{2}{\varepsilon}{}^\gamma-
  \stackrel{2}{\varepsilon}{}^\beta
  \stackrel{1}{\varepsilon}{}^\gamma
  F_{\beta\gamma}^{\tck\tck\alpha}\ ,              \\
\label{237}
&&\stackrel{3}{\omega}{}^{\tck\beta}_\alpha=
  -\stackrel{2}{\omega}{}^{\tck\gamma}_\alpha
  \stackrel{1}{\omega}{}^{\tck\beta}_\gamma+
  \stackrel{1}{\omega}{}^{\tck\gamma}_\alpha
  \stackrel{2}{\omega}{}^{\tck\beta}_\gamma-
  \stackrel{2}{\varepsilon}{}^\gamma
  \stackrel{1}{\varepsilon}{}^\delta
  F_{\gamma\delta\alpha}^{\tck\tck\beta}\ .
\eea

It is straightforward to show that the corresponding formulae for
$[\stackrel{2}{\Pi}\stackrel{1}{\Pi}]$ as a applied to the matter field
and the tetrad lead to the same parameters (\ref{236}), (\ref{237}).
These results are natural generalizations of the corresponding
commutator in an $M_4$. One can take (\ref{236}), (\ref{237}) as the
ultimate justification for attributing a fundamental significance to
the notion of a local P-transformation.

\subsection{Local Kinematical Inertial Frames}

As we have seen in (\ref{18}) , (\ref{19}), in an $M_4$ we can always
trivialize the gauge potentials globally. Since spacetime looks
minkowskian from a local point of view, it should be possible to
trivialize the gauge potentials in a $U_4$ locally, i.e.,
\be
\label{238}
\left\{
\begin{array}{rl}
e_i^{\tck\alpha}\big(x^k=\mathop{x}\limits_{0}\,^k\big)
\stackrel{\star}{=}\delta^\alpha_i\ ,\\[9pt]
\Gamma_i^{\tck\alpha\beta}\big(x^k=\mathop{x}\limits_{0}\,^k\big)
\stackrel{\star}{=}0.
\end{array} \right\}
\ee
The proof runs as follows: We rotate the tetrads according to
${e^\prime}^{\tck\alpha}_i=\Lambda_\beta^{\tck\alpha}
e_i^{\tck\beta}$. This induces a transformation of the connection,
namely the finite version of (\ref{227}) for $\varepsilon^\gamma=0$:
\be
\label{239}
{\Gamma^\prime}_{i\alpha}^{\tck\tck\beta}=
\Lambda_\alpha^{\tck\gamma}
\Lambda_\delta^{\tck\beta}
\Gamma_{i\gamma}^{\tck\tck\delta}-
\Lambda_\alpha^{\tck\gamma}\partial_i
\Lambda_\gamma^{\tck\beta}\ .
\ee
By a suitable choice of the rotation, we want these transformed
connection coefficients to vanish. We put (\ref{239}) provisionally
equal to zero, solve for $\partial_i
\Lambda_\beta^{\tck\alpha}$, and find
\be
\label{240}
\partial_i\Lambda_\beta^{\tck\alpha}=
\Lambda_\gamma^{\tck\alpha}
\Gamma_{i\beta}^{\tck\tck\gamma}
\Big(x^k=\mathop{x}\limits_{0}\,^k\Big)\ .
\ee
For prescribed $\Gamma_{i\beta}^{\tck\tck\gamma}$ at
$x^k=\mathop{x}\limits_{0}\,^k$ we can always solve this first order
linear differential equation, which concludes the first part
of the proof. Then we adjust the holonomic coordinates. The
connection $\Gamma_i^{\tck\alpha\beta}\stackrel{\star}{=}0$,
being a coordinate vector, stays zero, whereas the tetrad transforms
as follows:
${e^\prime}^{\tck\alpha}_i=(\partial x^k/\partial {x^\prime}^i)
e_k^{\tck\alpha}$. For a transformation of the type
${x^\prime}^i=\delta^i_\beta e_\ell^{\tck\beta}x^\ell$+ const. we find
indeed ${e^\prime}^{\tck\alpha}_i\stackrel{\star}{=}\delta_i^\alpha$, q.e.d.\symbolfootnote[1]{The proof was first given by von der Heyde \cite{34}, see also Meyer \cite{35}.}

What is the physical meaning of these trivial gauge frames existing all
over spacetime? Evidently they represent in a Riemann-Cartan spacetime
what was in Einstein's theory the freely falling non-rotating elevator.
For these considerations it is vital, however, that from our gauge
theoretical point of view the potentials
$(e_i^{\tck\alpha},\Gamma_i^{\tck\alpha\beta})$ are locally measurable,
whereas torsion and curvature, as derivatives of the potentials, are
only to be measured in a nonlocal way. For a local observer the world
looks minkowskian. If he wants to determine, e.g., whether his world
embodies torsion, he has to communicate with his neighbors thereby
implying nonlocality. This example shows that in the PG the question
whether spacetime carries torsion or not (or curvature or not) is not a
question one should ask one local observer.\symbolfootnote[2]
{Practically speaking, such non-local measurements may very well be
made by one observer only. Remember that, in the context of GR, the
Weber cylinder is also a non-local device for sensing curvature, i.e.
the cylinder is too extended for an Einstein elevator.}

It is to be expected that non-local quantities like torsion and
curvature, in analogy to Maxwell's theory and GR, are governed by field
equations. In other words, whether, for instance, the world is
riemannian or not, should in the framework of the PG not be imposed ad
hoc but rather left as a question to dynamics.

We call the frames (\ref{238}) ``local kinematical inertial frames" in
order to distinguish them from the local ``dynamical" inertial frames
in Einstein's GR. In the PG the notion of inertia refers to translation
and rotation, or to mass \emph{and} spin. A coordinate frame
$\gr{\partial}_i$ of GR has to fulfill the differential constraint
$\Omega_{ij}^{\tck\tck\alpha}(\gr{\partial}) \equiv0$, see eq.
(\ref{15}). Hence a tetrad frame, which is unconstrained, can move more
freely and is, as compared to the coordinate frame, a more local
object. Accordingly, the notion of inertia in the PG is more local than
that in GR. This is no surprise, since a test particle of GR carries a
mass $m$ which is a quantity won by integration over an extended
energy-momentum distribution. The matter field $\psi(x)$, however, the
object of consideration in the PG, is clearly a more localized being.

A natural extension of the Einstein equivalence principle to the PG
would then be to postulate that in the frames (\ref{238}) (these are
our new ``elevators") special relativity is valid locally. Consequently
the special-relativistic matter lagrangian $L$ in (\ref{jdeset}) should
in a $U_4$ be a lagrangian density ${\cal L}$ which couples to
spacetime according to
\be
\label{241}
{\cal L} ={\cal L}\Big(
\psi,\partial_i\psi, e_i^{\tck\alpha},\Gamma_i^{\tck\alpha\beta}
\Big)\stackrel{\star}{=}L(\psi,\partial_i\psi)\ ,
\ee
i.e. in the local kinematical inertial frames everything looks special
relativistic. Observe that derivatives of
$(e_i^{\tck\alpha},\Gamma_i^{\tck\alpha\beta})$ are excluded by our
``local equivalence principle".\symbolfootnote[1] {This principle was
formulated by von der Heyde \cite{34}, see also von der Heyde and Hehl
\cite{36}. In his seminar Dr. Rumpf \cite{17} has given a careful and
beautiful analysis of the equivalence principle in a Riemann-Cartan
spacetime. In particular the importance of his proof how to distinguish
the macroscopically indistinguishable teleparallelism and Riemann
spacetimes (see our Section 3.3) should be stressed.} Strictly this
discussion belongs into Lecture 3. But the geometry is so suggestive to
physical applications that we cannot resist the temptation to present
the local equivalence principle already in the context of spacetime
geometry.

Naturally, as argued above, the local equivalence principle is not to
be applied to directly observable objects like mass points, but rather
to the more abstract notion of a lagrangian. In a field theory there
seems to be no other reasonable option. And we have seen that the
fermionic nature of the building blocks of matter require a field
description, at least on a $c$-number level. Accordingly (\ref{241})
appears to be the natural extension of Einstein's equivalence principle
to the PG.

\subsection{Riemann-Cartan Spacetime Seen Anholonomically and
Holonomically}

We have started our geometrical game with the
$(e_i^{\tck\alpha},\Gamma_i^{\tck\alpha\beta})$-set. We would now like
to provide some machinery for translating this anholonomic formalism
into the holonomic formalism commonly more known at least under
relativists. Let us first collect some useful formulae for the
anholonomic regime. The determinant $e:= \textrm{det}\,
e_i^{\tck\alpha}$ is a scalar density, furthermore, by some algebra we
get $D_ie=\partial_ie$. If we apply the Leibniz rule the definitions
(\ref{25}) and (\ref{210}), we find successively $(F_\alpha:=
F_{\alpha\gamma}^{\tck\tck \gamma})$
\bea
\label{242}
&&D_i(ee^i_{\tck\alpha})=eF_\alpha\ ,                \\
\label{243}
&&2D_j\Big( ee^i_{\tck[\alpha}e^j_{\tck\beta]}\Big)=
  e\Big( F_{\alpha\beta}^{\tck\tck i}+2e^i_{\tck[\alpha}
  F_{\beta]}\Big)\ ,                                \\
\label{244}
&&D_{[\alpha}(e^i_{\tck\beta}e^j_{\tck\gamma]})=
  e^{[i}_{\,\tck[\alpha}F^{\tck\tck \ j]}_{\beta\gamma]}\ .
\eea
The last formula was convenient for rewriting the 2nd Bianchi identity
(\ref{216}), which was first given in a completely anholonomic form.
Eq. (\ref{243}), defining the ``modified torsion tensor" on its right
hand side, will be used in the context of the field equations to be
derived in Section 4.4.

Now, according to (\ref{dvajedan}), $dx^i\Gamma_i^{\tck\alpha\beta}
\til{e}_\beta$ is the relative rotation encountered by a tetrad
$\til{e}_\alpha$ in going from $x^k$ to $x^k+dx^k$. From this we can
calculate that the relative rotation of the respective coordinate frame
$\gr{\partial}_j=e_j^{\tck\alpha}\til{e}_\alpha$ is
$dx^i(\Gamma_{i\beta}^{\tck\tck\alpha}e_j^{\tck\beta}+
\partial_ie_j^{\tck\alpha})\til{e}_\alpha$. In a holonomic
coordinate system, the parallel transport is thus given by
\be
\label{245}
\nabla_i: =\partial_i+\tilde{\Gamma}_{ij}^{\tck\tck k}
h_k^{\tck j}\ ,
\ee
where $h$ represents the generator of coordinate transformation for
tensors fields and
\be
\label{246}
\tilde{\Gamma}_{ij}^{\tck\tck k}: =e_j^{\tck\alpha}
e^k_{\tck\beta}
\Gamma_{i\alpha}^{\tck\tck\beta}+
e^k_{\tck\beta}\partial_ie^{\tck\beta}_j\ .
\ee
This relation translates the anholonomic into the holonomic connection.
Observe that for a connection the conversion of holonomic to
anholonomic indices and vice versa is markedly different from the
simple transvection rule as applied to tensors. The holonomic
components of the covariant derivative of a tensor $A$ are given with
respect to its anholonomic components by
\be
\label{247}
\nabla_iA_{j\tck\tck}^{\ \ \ \ k\tck\tck}=
e_j^{\tck\alpha}\tck\tck \ e^k_{\tck\beta}
\tck\tck \  D_iA_{\alpha\tck\tck}
^{\ \ \ \ \beta\tck\tck}\ .
\ee

The concept of parallelism with respect to a coordinate frame, as
defined in (\ref{246}), is by construction locally identical with
minkowskian parallelism, as is measured in a local tetrad. In a similar
way, the local minkowskian length and angle measurements define the
metric in a coordinate frame:
\be
\label{248}
g_{ij}(x^k): =e_i^{\tck\alpha}(x^k)e_j^{\tck\beta}
(x^k)\eta_{\alpha\beta}\ .
\ee

From the antisymmetry $\Gamma_i^{\tck\alpha\beta}=
-\Gamma_i^{\tck\beta\alpha}$ of the anholonomic connection and from
(\ref{247}) results again $D_i\eta_{\alpha\beta}=0$, a relation which
we used already earlier, and
\be
\label{249}
\nabla_ig_{jk}=0\ ,
\ee
the so-called metric postulate of spacetime physics.

If we resolve (\ref{249}) with respect to
$\tilde{\Gamma}_{ij}^{\tck\tck k}$, we get
\be
\label{250}
\tilde{\Gamma}_{ij}^{\tck\tck k}=
\left\{
\begin{array}{rl}
k\\
ij
\end{array} \right\}+\frac{1}{2}F_{ij}^{\tck\tck k}
-\frac{1}{2}F_{j\tck i}^{\tck k}
+\frac{1}{2}F^k_{\tck ij}
\ee
and, taking (\ref{246}) into account, the corresponding relation
for the anholonomic connection:
\be
\label{251}
\Gamma_{\alpha\beta\gamma}: =
e^i_{\tck\alpha}\Gamma_{i\beta\gamma}=
(-\Omega_{\alpha\beta\gamma}+\Omega_{\beta\gamma\alpha}-
\Omega_{\gamma\alpha\beta}+F_{\alpha\beta\gamma}-
F_{\beta\gamma\alpha}+F_{\gamma\alpha\beta})/2\ .
\ee
We have introduced here the Christoffel symbol $\left\{
\begin{array}{rl}k\\ ij
\end{array} \right\}$, the holonomic components of the
torsion tensor,
\be
\label{252}
F_{ij}^{\tck\tck k}=e^k_{\tck\alpha}
F_{ij}^{\tck\tck\alpha}=
2\tilde{\Gamma}_{[ij]}^{\,\tck\tck\, k}\ ,
\ee
and the object of anholonomity
\be
\label{253}
\Omega_{\alpha\beta}^{\tck\tck\gamma}: =
e^i_{\tck\alpha}e^j_{\tck\beta}
\Omega_{ij}^{\tck\tck\gamma}\ ;\quad
\Omega_{ij}^{\tck\tck\gamma}: =
2\partial_{[i}e_{j]}^{\tck\,\gamma}\ .
\ee
Expressing the holonomic components of the curvature tensor in
terms of $\tilde{\Gamma}_{ij}^{\tck\tck k}$ yields
\be
\label{254}
F_{ij\,k}^{\tck\tck\tck\ell}=2\Big(
\partial_{[i}\tilde{\Gamma}_{j]k}^{\tck \,\tck\ell}+
\tilde{\Gamma}_{[i|m}^{\,\tck\,\tck\ell}
\tilde{\Gamma}_{|j]k}^{\,\tck \,\tck m}\Big)\ .
\ee
Finally, taking the antisymmetric part of (\ref{246}) or using the
definition of torsion (\ref{210}), we get
\be
\label{255}
F_{ij}^{\tck\tck\alpha}=
\Omega_{ij}^{\tck\tck\alpha}-
2e_{[i}^{\,\tck\beta}
\Gamma_{j]\beta}^{\tck\,\tck\alpha}\ ,
\ee
a formula which will play a key role in discussing macroscopic gravity:
the object of anholonomity mediates between torsion and the anholonomic
connection.

Eq. (\ref{250}) shows that instead of the potentials
$(e_i^{\tck\alpha},\Gamma_i^{\tck\alpha\beta})$ we can use
holonomically the set\symbolfootnote[1]{Note added in 2023: We had identified Schouten's [3] notation of the torsion tensor $S_{ij}{}^k$ with our $F_{ij}{}^k$:
   $F_{ij}{}^k \equiv S_{ij}{}^k$.}
$(g_{ij}, S_{ij}^{\tck\tck k})$, the geometry is
always a Riemann-Cartan one, only the mode of description is different.
In the holonomic description $e_i^{\tck\alpha}$ enters the definition
of torsion $F_{ij}^{\tck\tck\alpha}$, hence we should not use
$F_{ij}^{\tck\tck\alpha}$ as an independent variable in place of
$\Gamma_i^{\tck\alpha\beta}$. The anholonomic formalism is superior in
a gauge approach, because
$(e_i^{\tck\alpha},\Gamma_i^{\tck\alpha\beta})$ are supposed to be
directly measurable and have an interpretation as potentials and the
corresponding transformation behavior, whereas the holonomic set
 $(g_{ij}, S_{ij}^{\tck\tck k})$ is a tensorial one.

\begin{figure}[htb]
\centering\vspace{-.4cm}
\includegraphics[height=4cm]{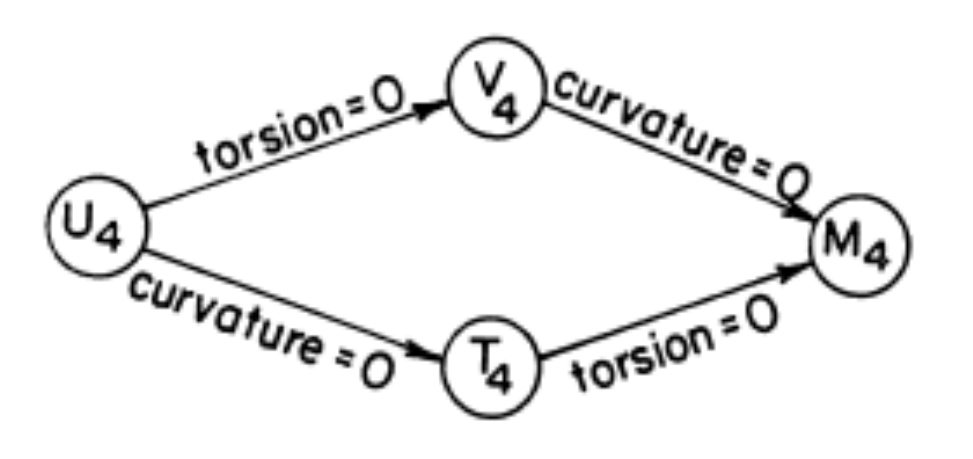}\vspace{-.4cm}
  \caption{The Riemann-Cartan spacetime $U_4$ and its limiting
  cases:\newline
  $T_4$ = teleparallelism, $V_4$ = Riemann, $M_4$ = Minkowski.}
\end{figure}

If we put curvature to zero, we get a spacetime with teleparallelism
$T_4$, a $U_4$ for vanishing torsion is a Riemann spacetime $V_4$, see
Figure 3.

Perhaps surprisingly, as regards to their physical degrees of freedom,
the $T_4$ and the $V_4$ are in some sense similar to each other. In a
$T_4$ the curvature vanishes, i.e. the parallel transfer is integrable.
Then we can always pick preferred tetrad frames such that the
connection vanishes globally:
\be
\label{256}
T_4:\ =\Big(e_i^{\tck\alpha}, \Gamma_i^{\tck\alpha\beta}
\stackrel{\star}{=}0\Big)\ ;\quad
\Big(
F_{ij}^{\tck\tck\alpha}
\stackrel{\star}{=}
\Omega_{ij}^{\tck\tck\alpha},
F_{ij}^{\tck\tck\alpha\beta}\equiv 0\Big)\ .
\ee
In these special anholonomic coordinates we are only left with the
tetrad $e_i^{\tck\alpha}$ as variable. In a $V_4$ the torsion vanishes,
i.e., according to (\ref{251}), the connection can be expressed
exclusively in terms of the (orthonormal) tetrads:
\bea
\label{257}
&&V_4: =\Big( e_i^{\tck\alpha},\Gamma_{\alpha\beta\gamma}=
-\frac{1}{2}\Omega_{\alpha\beta\gamma}+\cdots\Big)\ ; \\
&&\Big(F_{ij}^{\tck\tck\alpha}\equiv 0,\quad
F_{ij\alpha}^{\tck\tck\tck\beta}=\partial_i
\Omega_{j\alpha}^{\tck\tck\beta}-\cdots\Big)\ .\nonumber
\eea
Again nothing but tetrads are left. In other words, in a $T_4$ as well
as in a $V_4$ the gauge variables left over are the tetrad coefficients
alone. We have to keep this in mind in discussing macroscopic gravity.

\subsection{Global P-Transformation and the \texorpdfstring{\mb{M_4}}{M\_4}}

In an appropriate P-gauge approach one should recover the global
P-trans\-for\-mation provided the $U_4$ degenerates to an $M_4$. In an
$M_4$, because of (\ref{256}), we can introduce a global tetrad system
such that $\Gamma_i^{\tck\alpha\beta} \stackrel{\star}{=}0$.
Furthermore the torsion vanishes, $F_{ij}^{\tck\tck\alpha}
\stackrel{\star}{=}2\partial_{[i}e_{j]}^{\tck\alpha}=0$, i.e. the
orthonormal tetrads, in the system with $\Gamma_i^{\tck\alpha\beta}
\stackrel{\star}{=}0$, represent holonomic frames of the cartesian
coordinates. This means that the conditions (\ref{238}) are now valid
on a \emph{global} level. If one only allows for P-transformations
linking two such coordinate systems, then the potentials don't change
under P-transformations and (\ref{226}), (\ref{227}) yield
\be
\label{258}
\left\{
\begin{array}{rl}
-\partial_i\varepsilon^\alpha+\omega_\gamma
^{\tck\alpha}\delta_i^{\tck\gamma}& =0\\[6pt]
-\partial_i\omega^{\alpha\beta}& =0
\end{array} \right\}
\ee
or
\be
\label{259}
\left\{
\begin{array}{rl}
\varepsilon^\alpha &=  \ \stackrel{0}{\varepsilon}{}^\alpha+
\stackrel{0}{\omega}{}_\gamma^{\tck\alpha}x^\gamma \\[4pt]
\omega^{\alpha\beta} &=  \ \stackrel{0}{\omega}{}^{\alpha\beta}
\end{array} \right\}
\ee
with the constants $\stackrel{0}{\varepsilon}{}^\alpha$ and
$\stackrel{0}{\omega}{}^{\alpha\beta}$. Substitution into (\ref{26})
leads to the global P-transformation of the matter field
\be
\label{260}
(\Pi\psi)(x)=[1-\stackrel{0}{\varepsilon}{}^\alpha\partial_\alpha
+\stackrel{0}{\omega}{}^{\alpha\beta}
(x_{[\beta}\partial_{\alpha]}+f_{\beta\alpha})]
\psi(x)\ .
\ee
In terms of the ``P-transformed" coordinates
\be
\label{261}
{x^\prime}^\gamma=x^\gamma-
\stackrel{0}{\omega}{}_\alpha^{\tck\gamma}x^\alpha-
\stackrel{0}{\varepsilon}{}^\gamma\ ,
\ee
it can be recast into the perhaps more familiar form
\be
\label{262}
(\Pi\psi)(x^\prime)=\Big(1+\stackrel{0}{\omega}{}^{\alpha\beta}
f_{\beta\alpha}\Big)\psi(x)\ .
\ee
Thus, in an $M_4$, the local P-transformation degenerates into the
global P-transformation, as it is supposed to do.

We have found in this lecture that spacetime ought to be described by a
Riemann-Cartan geometry, the geometrical gauge variables being the
potentials $(e_i^{\tck\alpha}, \Gamma_i^{\tck\alpha\beta})$. The main
results are, amongst other things, collected in the table of Section
3.6.

\lecture{Lecture 3: Coupling of Matter to Spacetime and the Two
General Gauge Field Equations}
\addcontentsline{toc}{section}{Lecture 3:~ Coupling of Matter
and the Two General Gauge Field Equations}

\subsection{Matter Lagrangian in a \texorpdfstring{\mb{U_4}}{U\_4} }
\label{matter}

In the last lecture we concentrated on working out the geometry of
spacetime. But already in (\ref{241}) we saw how to extend reasonably
the Einstein equivalence principle to the ``local equivalence
principle" applicable to the PG. Hence we postulate the action function
of matter as coupled to the geometry of spacetime to read
\be
\label{31}
W_m=\int d^4x{\cal L}\Big[\eta_{\alpha\beta}, \gamma^\alpha\cdots,
\psi(x), \partial_i\psi(x),e_i^{\tck\alpha}(x),
\Gamma_i^{\tck\alpha\beta}(x)\Big]\ .
\ee
Observe that the local Minkowski metric $\eta_{\alpha\beta}$, the Dirac
matrices $\gamma^\alpha\cdots$ etc., since referred to the tetrads,
maintain their special relativistic values in (\ref{31}).

In a local kinematical inertial frame (\ref{238}), the potentials
$(e_i^{\tck\alpha}, \Gamma_i^{\tck\alpha\beta})$ can be made trivial
and, in the case of vanishing torsion and curvature, this can be done
even globally, and then we fall back to the special-relativistic action
function (\ref{jdeset}) we started with.

The lagrangian in (\ref{31}) is of first order by assumption, i.e. only
first derivatives of $\psi(x)$ enter. If we would allow for higher
derivatives, we could not be sure of how to couple to
$(e_i^{\tck\alpha},\Gamma_i^{\tck\alpha\beta})$: the higher derivatives
would presumably ``feel" not only the potentials, but also the
non-local quantities torsion and curvature. In such a case the gauge
field strengths themselves would couple to $\psi(x)$ and thereby break
the separation between matter and gauge field lagrangian. Then we would
lose the special-relativistic limit seemingly necessary for executing a
successful P-gauge approach.

Our postulate (\ref{31}) applies to matter fields $\psi(x)$. These
fields are anholonomic objects by definition. Gauge potentials of
internal symmetries, like the electromagnetic potential $A_i(x)$,
emerging as holonomic covariant vectors (one-forms), must not couple to
$(e_i^{\tck\alpha}, \Gamma_i^{\tck\alpha\beta})$. Otherwise gauge
invariance, in the case of $A_i$ the $U(1)$-invariance, would be
violated. This implies that gauge bosons other than the
$(e_i^{\tck\alpha}, \Gamma_i^{\tck\alpha\beta})$-set, and in particular
the photon field $A_i$, will be treated as P-scalars. Because of the
natural division of physical fields into matter fields and gauge
potentials (see Section 1.3), we cannot see any disharmony in exempting
the internal gauge bosons from the coupling to the $(e_i^{\tck\alpha},
\Gamma_i^{\tck\alpha\beta})$-set.\symbolfootnote[1] {There are opposing
views, however, see Hojman, Rosenbaum, Ryan, and Shepley \cite{37,38}.
The tlaplon concept is a possibility to circumvent our arguments, even
if not a very natural one, as it seems to us. According to Ni
\cite{39}, the tlaplon theories are excluded by experiment. See also
Mukku and Sayed \cite{40}.}

Besides the matter field $\psi(x)$, the $(4+6)$ P-gauge potentials are
new independent variables in (\ref{31}). By means of the action
principle we can derive the matter field equation. Varying (\ref{31})
with respect to $\psi(x)$ yields\symbolfootnote[1] {The variational
derivative of a function
$f=f(\psi,\partial_i\psi,\partial_i\partial_k\psi,\cdots)$ is defined
by $\frac{\delta f}{\delta\psi}:=\frac{\partial f}{\partial\psi}
-\partial_i\frac{\partial f}{\partial\partial_i\psi}
+\partial_i\partial_k\frac{\partial f}
{\partial\partial_i\partial_k\psi}-+\cdots $.}
\be
\label{32}
\delta{\cal L}/\delta\psi = 0\ .
\ee
In our subsequent considerations we'll always assume that (\ref{32}) is
fulfilled.

\subsection{Noether Identities~of~the Matter Lagrangian:
\texorpdfstring{\\}{}
Identification of Currents and Conservation Laws}
\label{noether}

The material action function (\ref{31}) is a P-scalar by construction.
Consequently it is invariant under active local P-transformations
$\Pi(x)$. The next step will then consist in exploiting this invariance
property of the action function according to the Noether procedure.

Let us call the field variables in (\ref{31}) collectively
$Q\in(\psi, e_i^{\tck\alpha},
\Gamma_i^{\tck\alpha\beta})$. Then the P-invariance demands
\be
\label{33}
\delta W_m:\ \int_{\Pi\Omega} d^4x{\cal L}(\Pi Q,\partial_i\Pi Q)
-\int d^4x{\cal L}(Q,\partial_iQ)\equiv 0\ ,
\ee
where $\Pi\Omega$ is the volume translated by an amount $\varepsilon
^\alpha(x)$. By the chain rule and by the Gauss law we calculate
\bea
\label{34}
&&\int_\Omega d^4x\Bigg[
\frac{\partial{\cal L}}{\partial Q}\delta Q+
\frac{\partial{\cal L}}{\partial\partial_iQ}
\delta\partial_iQ
\Bigg]+\int_{\partial\Omega}
dA_i\varepsilon^i{\cal L}\nonumber\\
&&=
\int_\Omega d^4x\Bigg[
\frac{\delta{\cal L}}{\delta Q}\delta Q+\partial_i
\Big(
\varepsilon^i{\cal L}+
\frac{\partial{\cal L}}{\partial\partial_iQ}
\delta Q\Big)\Bigg]\equiv0\ .
\eea
Since this expression is valid for an arbitrary volume $\Omega$, the
integrand itself has to vanish. Furthermore we can substitute
$\partial_i\to D_i$ in (\ref{34}), since the expression in the
parenthesis carries no anholonomic indices:
\be
\label{35}
\frac{\delta{\cal L}}{\delta Q}\delta Q+D_i
\Big(
\varepsilon^i{\cal L}+
\frac{\partial{\cal L}}{\partial\partial_iQ}
\delta Q\Big)\equiv 0\ .
\ee
The identity (\ref{35}) is valid quite generally for any lagrangian
${\cal L}(Q,\partial_iQ)$. We will need it later-on also for discussing
the properties of the gauge field lagrangian.

Going back to (\ref{31}), we find using (\ref{32}),
\be
\label{36}
\frac{\delta{\cal L}}{\delta e_i^{\tck\alpha}}
\delta e_i^{\tck\alpha}+
\frac{\delta{\cal L}}{\delta\Gamma_i^{\tck\alpha\beta}}
\delta\Gamma_i^{\tck\alpha\beta}+D_i
\Big(\varepsilon^\alpha
e^i_{\tck\alpha}{\cal L}+
\frac{\partial{\cal L}}{\partial\partial_i\psi}\delta\psi
\Big)\equiv0\ .
\ee
Now we substitute the P-variations (\ref{226}), (\ref{227}), (\ref{26})
of $(e_i^{\tck\alpha}, \Gamma_i^{\tck\alpha\beta})$ and of $\psi$,
respectively, into (\ref{36}):
\bea
\label{37}
\frac{\delta{\cal L}}{\delta e_i^{\tck\alpha}}\Big(
-D_i\varepsilon^\alpha+\omega_\gamma^{\tck\alpha}
e_i^{\tck\gamma}-
\varepsilon^\gamma F_{\gamma i}^{\tck\tck\alpha}
\Big)+
\frac{\delta{\cal L}}{\partial\Gamma_i^{\tck\alpha\beta}}
\Big(
-D_i\omega^{\alpha\beta}-
\varepsilon^\gamma F_{\gamma i}^{\tck\tck\alpha\beta}
\Big)+\nonumber\\
+
D_i
\Bigg[\varepsilon^\alpha
e^i_{\tck\alpha}{\cal L}+
\frac{\partial{\cal L}}{\partial\partial_i\psi}\Big(
-\varepsilon^\gamma D_\gamma +\omega^{\alpha\beta} f_{\beta\alpha}
\Big)\psi\Bigg]\equiv 0\ .
\eea
We differentiate the last bracket and order according to the
independent quantities $D_i\varepsilon^\alpha$,
$D_i\omega^{\alpha\beta}$, $\varepsilon^\alpha$,
$\omega^{\alpha\beta}$, the coefficients of which have to vanish
separately. This yields th $(10+40)$ identities
\bea
\label{38}
&& e\sum_\alpha^{\tck i}:=\
\frac{\delta{\cal L}}{\delta e_i^{\tck\alpha}}\equiv
e^i_{\tck\alpha}{\cal L}-
\frac{\partial{\cal L}}{\partial\partial_i\psi}D_\alpha\psi\ ,\\
\label{39}
&& e\tau_{\alpha\beta}^{\tck\tck i}:=\
\frac{\delta{\cal L}}{\delta \Gamma_i^{\tck\alpha\beta}}\equiv
-\frac{\partial{\cal L}}{\partial\partial_i\psi}f_{\alpha\beta}\psi\,\\
\label{310}
&& D_i\Big(e\Sigma_\alpha^{\tck i}\Big)\equiv
F_{\alpha i}^{\tck\tck\beta}e\Sigma_\beta^{\tck i}
+ F_{\alpha i}^{\tck\tck\beta\gamma}e\tau_{\beta\gamma}
^{\tck\tck i}\ ,            \\
\label{311}
&& D_i\Big(e\tau_{\alpha\beta}^{\tck\tck i}\Big)-
e\Sigma_{[\alpha\beta]}\equiv0\ .
\eea

Our considerations are valid for any $U_4$, especially for an $M_4$. In
an $M_4$ in the global coordinates (\ref{18}), (\ref{19}), the
equations (\ref{310}), (\ref{311}) degenerate to the
special-relativistic momentum and angular momentum conservation laws,
respectively:
\bea
\label{312}
&&\partial_\gamma\Sigma_\alpha^{\tck\gamma}=0\ , \\[3pt]
\label{313}
&&\partial_\gamma\tau_{\alpha\beta}^{\tck\tck\gamma}-
\Sigma_{[\alpha\beta]}=0\ .
\eea
Consequently $\Sigma_\alpha^{\tck i}$ is the canonical momentum
current, linked via (\ref{38}) to the translational potential
$e_i^{\tck\alpha}$, and $\tau_{\alpha\beta}^{\tck\tck i}$ is the
canonical spin current, linked via (\ref{39}) to the rotational
potential $\Gamma_i^{\tck\alpha\beta}$. Moreover, (\ref{310}),
(\ref{311}) are recognized as the momentum and angular momentum
conservation laws in a $U_4$.

It comes to no surprise that in $U_4$ the volume-force densities of the
Lorentz type $F_{\alpha i}^{\tck\tck\beta}(e\Sigma_\beta^{\tck i})$ and
$F_{\alpha i}^{\tck\tck\beta\gamma}(e\tau_{\beta\gamma}^ {\tck\tck
i})$, respectively, appear on the right hand side of the momentum
conservation law. The analog of the latter force is known in GR as the
Matthisson force acting on a spinning particle,\symbolfootnote[1] {In
\cite{41} we compared in some detail the standing of the Matthisson
force in GR with that in the $U_4$-framework. Clearly the Matthisson
force emerges much more natural in the PG.} and because of the similar
couplings of translations and rotations, the force $F_{\alpha
i}^{\tck\tck\beta}(e\Sigma_\beta^{\tck i})$ is to be expected, too. The
left hand side of (\ref{310}) contains second derivatives of the matter
field $\psi(x)$. It is because of this ``non-locality" that the local
equivalence principle doesn't apply on this level. Hence the volume
forces just discussed, do not violate the local equivalence principle.

One further observation in the context of the Noether identities
(\ref{38}), (\ref{39}) is of importance. Because of
\be
\label{314}
-\frac{\partial{\cal L}}{\partial\partial_i\psi}f_{\alpha\beta}\psi
=\frac{\partial{\cal L}}{\partial\partial_k\psi}
\frac{\partial(D_k\psi)}{\partial\Gamma_i^{\tck\alpha\beta}}
\stackrel{!}{=}
\frac{\partial{\cal L}}{\partial\Gamma_i^{\tck\alpha\beta}}\ ,
\ee
the connection must only show up in the lagrangian in terms of
$D_i\psi$. Similarly, $e_i^{\tck\alpha}$ can only enter as
$e=\textrm{det}\,e^{\tck\alpha}_i$ and in transvecting $D_i\psi$
according to the substitution (cf.(\ref{jdeset}))
\be
\label{315}
L(\psi,\partial_i\psi)\to eL(\psi,e_{\tck\alpha}^iD_i\psi)=
{\cal L}(\psi, D_\alpha\psi,e)\ ,
\ee
because then
\be
\label{316}
\frac{\partial{\cal L}}{\partial e_i^{\tck\alpha}}=
\frac{\partial{\cal L}}{\partial e}
\frac{\partial e}{\partial e_i^{\tck\alpha}}+
\frac{\partial{\cal L}}{\partial D_\beta\psi}
\frac{\partial(D_\beta\psi)}{\partial e_i^{\tck\alpha}}=
e\Sigma_\alpha^{\tck i}\ ,
\ee
q.e.d. Therefore the so-called \emph{minimal coupling} (\ref{315}) is a
consequence of (\ref{241}), (\ref{31}) and of local P-invariance. It is
derived from the local equivalence principle.\symbolfootnote[2]{There
have been several attempts to develop non-minimal coupling procedures,
see Cho \cite{42}, for instance.}

The main results of the kinematical considerations of Sections
\ref{matter} and \ref{noether} are again collected in the table of
Section 3.6

\subsection{The Degenerate Case of Macroscopic (Scalar) Matter}

From GR we know that the equations of motion for a test particle moving
in a given field are derived by integrating the momentum conservation
law. This will be similar in the PG. However, one has to take into
account the angular momentum conservation law additionally.

A test particle in GR, as a macroscopic body, will consist of many
elementary particles. Hence in order to derive its properties from
those of the elementary particles, one has, in the sense of a
statistical description, to average over the ensemble of particles
constituting the test body.

Mass is of a monopole type and adds up, whereas spin is of a dipole
type and normally tends to be averaged out (unless some force aligns
the spins like in ferromagnets or in certain superfluids).

Accordingly macroscopic matter, and in particular the test particles of
GR, will carry a finite energy-momentum whereas the spin is averaged
out, i.e. $\langle\tau_{\alpha\beta}^{\tck \tck i}\rangle\sim0$.
Consequently, because of the macroscopic analog of (\ref{311}), the
macroscopic energy-momentum tensor $eT_\alpha^{\tck i}=\langle
e\Sigma_\alpha^{\tck i}\rangle$ turns out to be symmetric, as we are
used to it in GR. What effect will this averaging have on the momentum
conservation law (\ref{310})? Provided the curvature doesn't depend
algebraically on the spin, the Matthisson force is averaged out and we
expect the macroscopic analog of (\ref{310}) to look like
\be
\label{317}
D_i\Big(eT_\alpha^{\tck i}\Big)\sim \langle F^{\tck\tck\beta}
_{\alpha i}\rangle eT_\beta^{\tck i}\ .
\ee
These arguments are, of course, not rigorous. But we feel justified in
modeling macroscopic matter by a scalar, i.e. a spinless field
$\Phi(x)$. And for a scalar field $\Phi(x)$ our derivations will become
rigorous. We lose thereby the information that macroscopic matter
basically is built up from fermions and should keep this in mind in
case we run into difficulties.

Let us then consider $U_4$-spacetime with only scalar matter
present.\symbolfootnote[1]{Compare for these considerations always the
lecture of Nitsch \cite{16} and the diploma thesis of Meyer \cite{35}
and references given there.} The lagrangian of the scalar field
$\Phi(x)$ reads
\be
\label{318}
{\cal L}=eL\Big(
\Phi, e^i_{\tck\alpha}\partial_i\Phi\Big)=
{\cal L} (\Phi,\partial_\alpha\Phi,e)\ .
\ee
If we denote the momentum current by $\sigma_\alpha^{\tck i}$,
we find via (\ref{39}), (\ref{311})
$\tau_{\alpha\beta}^{\tck\tck i}=0$ and
$\sigma_{[\alpha\beta]}=0$. Thus (\ref{310}) yields
\be
\label{319}
D_i\Big(e\sigma_\alpha^{\tck i}\Big)-
F_{\alpha i}^{\tck\tck\beta}e\sigma_\beta^{\tck i}=0\ ,
\ee
and only one type of volume-force density is left.

Because of the symmetry of $\sigma_{\alpha\beta}$, the term
$\sim\Gamma_{\alpha\gamma}^{\tck\tck\beta}\sigma_\beta ^{\tck\gamma}$
contained in the volume force density vanishes identically and we find
\be
\label{320}
\partial_i\Big(e\sigma_\alpha^{\tck i}\Big)-
\Omega_{\alpha i}^{\tck\tck\beta}e\sigma_\beta^{\tck i}=0\ ,
\ee
or, after some algebra,
\be
\label{321}
\partial_i\Big(e\sigma_\alpha^{\tck i}\Big)
 -\left\{\begin{array}{rl}i\\ ik
\end{array} \right\}e\sigma_\alpha^{\tck k}=
\stackrel{\{\}}{\nabla_i}\Big(e\sigma_\alpha^{\tck i}\Big)=0\ ,
\ee
i.e. the \emph{rotational potential} $\Gamma_i^{\tck\alpha\beta}$
\emph{drops out} from the momentum law of a \emph{scalar} field
altogether. The covariant derivative in (\ref{321}) is understood with
respect to the Christoffel symbol. We stress that the volume force of
(\ref{319}) is no longer manifest, it got ``absorbed". Consequently a
scalar field $\Phi(x)$ is not sensitive to the connection
$\Gamma_i^{\tck\alpha\beta}$. It is perhaps remarkable that this
property of $\Phi(x)$ is a result of the Noether identities as applied
to an arbitrary scalar matter lagrangian, i.e. we need no information
about the gauge field part of the lagrangian in order to arrive at
(\ref{320}) and (\ref{321}), respectively. It is a universal property
of any scalar matter field embedded in a general $U_4$.

For the Maxwell potential $A_i$, which is treated in the PG as a scalar\\
(-valued one-form), all these considerations apply mutatis mutandi. It
should be understood, however, that spinning matter, say Dirac matter,
couples to $\Gamma_i^{\tck\alpha\beta}$, and in this case there is no
ambiguity left as to whether we live in a $V_4$, a $T_4$ or in a
general $U_4$. Therefore a Dirac \emph{electron} can be used as a probe
for measuring the rotational potential\symbolfootnote[1] {The equations
of motion of a Dirac electron in a $U_4$, and in particular its
precession in such a spacetime, were studied in detail by Rumpf
\cite{17}. For earlier reference see \cite{1}. Recent work includes
Hojman \cite{43}, Balachandran et al. \cite{44} and the extensive
studies of Yasskin and Stoeger \cite{20,45,46}.}
$\Gamma_i^{\tck\alpha\beta}$.

Let us conclude with some plausibility considerations: In our model
universe filled only with scalar matter, $\Phi(x)$ does not feel,
$\Gamma_i^{\tck\alpha\beta}$ as we saw. Hence one should expect that it
doesn't produce it either, or, in other words, the ``scalar" universe
should obey a teleparallelism geometry $T_4$ with the rigid
$F_{ij}^{\tck\tck\alpha\beta}=0$ constraint, since then, according to
(\ref{256}), we could make $\Gamma_i^{\tck\alpha\beta}$ vanish
globally. Because of the equivalence of (\ref{319}), (\ref{320}), and
(\ref{321}), scalar matter would move along geodesics of the attached
$V_4$, nevertheless. If one took care that the field equations of the
$T_4$ were appropriately chosen, one could produce a $T_4$-theory which
is, for scalar matter, indistinguishable from GR.

To similar conclusions leads the following argument: Suppose there
existed only scalar matter. Then there is no point in gauging the
rotations since $\Phi(x)$ is insensitive to it. Repeating all
considerations of Lecture 2, yields immediately a $T_4$ as the
spacetime appropriate for a translational gauge theory, in consistence
with the arguments as given above.

Summing up: scalar (macroscopic) matter is uncoupled from the
rotational potential $\Gamma_i^{\tck\alpha\beta}$ , as proven in
(\ref{320}), and is expected to span a $T_4$-spacetime.

\subsection{General Gauge Field Lagrangian and its
Noether Identitites}

In order to build up the total action function of matter plus field,
one has to add to the matter Lagrangian in (\ref{31}) a gauge field
lagrangian $V$ representing the effect of the free gauge field. We will
assume, in analogy to the matter lagrangian, that the gauge field
lagrangian is of first order in the gauge potentials:
\be
\label{322}
V=V\Big(
\kappa_1,\kappa_2\cdots,\eta_{\alpha\beta},
e_i^{\tck\alpha},\Gamma_i^{\tck\alpha\beta},\partial_ke_i^{\tck\alpha}
,\partial_k \Gamma_i^{\tck\alpha\beta}\Big)\ .
\ee
The quantities $\kappa_1,\kappa_2\dots$ denote some universal coupling
constants to be specified later and for parity reasons we assume that
$V$ must not depend on the Levi-Civit$\grave{\textrm{a}}$ symbol
$\varepsilon^{\alpha \beta\gamma\delta}$. Then the gauge field
equations, in analogy to Maxwell's theory, will turn out to be of
second order in the potentials in general.

Applying the Noether identity (\ref{35}) to
(\ref{322})\symbolfootnote[1] {The identities of this section and of
Section 3.2 can be also found in the paper of W. Szczyrba \cite{18}.}
yields
\be
\label{323}
\frac{\delta V}{\delta e_i^{\tck\alpha}}\delta e_i^{\tck\alpha}
+\frac{\delta V}{\delta \Gamma_i^{\tck\alpha\beta}}\delta
\Gamma_i^{\tck\alpha\beta}+D_j\Big(
\varepsilon^jV+{\cal H}_\alpha^{\tck ij}
\delta e_i^{\tck\alpha}+{\cal H}_{\alpha\beta}^{\,\tck\tck ij}
\delta \Gamma_i^{\tck\alpha\beta}\Big)\equiv 0\ ,
\ee
where we have introduced the field momenta\symbolfootnote[2] {In spite
of current practice in theoretical physics, it should be stressed that
even in microphysical vacuum electrodynamics it is advisable to
introduce the ``induction" tensor density ${\cal H}^{ij}$ as an
independent concept amenable to direct operational interpretation (see
Post \cite{47}). One is in good company then (Maxwell). For a
``practical" application of such ideas see the discussion preceding eq.
(\ref{455}). Recent work of Rund \cite{48} seems to indicate that also
in Yang-Mills theories such a distinction between induction and field
could be useful.}
\be
\label{324}
{\cal H}_\alpha^{\tck ij}:= \frac{\partial V}
{\partial\partial_je_i^{\tck\alpha}}\ ,\qquad
{\cal H}_{\alpha\beta}^{\,\tck\tck ij}:= \frac{\partial V}
{\partial\partial_j\Gamma_i^{\tck\alpha\beta}}\ ,
\ee
canonically conjugated to the potentials $e_i^{\tck\alpha}$ and
$\Gamma_i^{\tck\alpha\beta}$, respectively. Substitute
(\ref{226}), (\ref{227}) into (\ref{323}) and get
\bea
\label{325}
&&\frac{\delta V}{\delta e_i^{\tck\alpha}}\Big(
-D_i\varepsilon^\alpha+\omega_\gamma^{\tck\alpha}
e_i^{\tck\gamma}-\varepsilon^\gamma F_{\gamma i}^{\tck\tck\alpha}
\Big)+
\frac{\delta V}{\delta\Gamma_i^{\tck\alpha\beta}}\Big(
-D_i\omega^{\alpha\beta}-\varepsilon^\gamma F_{\gamma i}^
{\tck\tck\alpha\beta}
\Big)+\nonumber\\
&&\hspace{30pt}+D_j\Big[
e^j_{\tck\alpha}\varepsilon^\alpha V+{\cal H}_\alpha
^{\tck ij}\Big(
-D_i\varepsilon^\alpha+\omega_\gamma^{\tck\alpha}
e_i^{\tck\gamma}-\varepsilon^\gamma F_{\gamma i}^{\tck\tck\alpha}
\Big)+ \nonumber\\
&&\hspace{30pt}+{\cal H}_{\alpha\beta}^{\tck\tck ij}\Big(
-D_i\omega^{\alpha\beta}-\varepsilon^\gamma F_{\gamma i}^
{\tck\tck\alpha\beta}\Big)\Big]\equiv0\ .
\eea
Again we have to differentiate the bracket. This time, however, we
find second derivatives of the translation and rotation parameters,
namely, collecting these terms,
\bea
\label{326}
&-&{\cal H}_\alpha^{\tck ij}D_jD_i\varepsilon^\alpha-
{\cal H}_{\alpha\beta}^{\,\tck\tck ij}D_jD_i\omega^{\alpha\beta}=
\nonumber\\
&&-
{\cal H}_\alpha^{\tck ij}D_{(i}D_{j)}\varepsilon^\alpha-
{\cal H}_{\alpha\beta}^{\,\tck\tck ij}D_{(i}D_{j)}\omega^{\alpha\beta}
+\nonumber\\
&&+\frac{1}{2}{\cal H}_\alpha^{\tck ij}F_{ij\gamma}^{\tck\tck
\tck\alpha}\varepsilon^\gamma+
{\cal H}_{\alpha\beta}^{\,\tck\tck ij}F_{ij\gamma}^{\tck\tck
\tck\alpha}\omega^{\gamma\beta}\ ,
\eea
where we have used (\ref{28}). Since there are no other second
derivative terms in (\ref{325}) but the ones which show up in
(\ref{326}), the coefficients of $D_{(i}D_{j)}\varepsilon^\alpha$
and $D_{(i}D_{j)}\omega^{\alpha\beta}$ in (\ref{326}) have to
vanish identically, i.e.
\be
\label{327}
{\cal H}_\alpha^{\tck (ij)}\equiv 0\ ,\qquad
{\cal H}_{\alpha\beta}^{\,\tck\tck (ij)}\equiv 0
\ee
or
\be
\label{328}
\frac{\partial V}{\partial\partial_{(j}e_{i)}^{\tck\alpha}}\equiv 0\ ,
\qquad
\frac{\partial V}{\partial\partial_{(j}\Gamma_{i)}^{\tck\alpha\beta}}
\equiv 0\ .
\ee
Accordingly, the derivatives $\partial_je_i^{\tck\alpha}$ and
$\partial_j\Gamma_i^{\tck\alpha\beta}$ can only enter $V$ in the form
$\partial_{[j}e_{i]}^{\tck\alpha}$ and
$\partial_{[j}\Gamma_{i]}^{\tck\alpha\beta}$, i.e. in the form present
in torsion and curvature. Algebraically one cannot construct out of
$\Gamma_i^{\tck\alpha\beta}$ a tensor piece for $V$ because of
(\ref{238}). Hence, using (\ref{324}), we have

\be
\label{329}
{\cal H}_\alpha^{\tck ij}=2\frac{\partial V}
{\partial F_{ij}^{\tck\tck\alpha}}\ ,\qquad
{\cal H}_{\alpha\beta}^{\,\tck\tck ij}=2\frac{\partial V}
{\partial F_{ij}^{\tck\tck\alpha\beta}}
\ee
or
\be
\label{330}
V=V\Big(\kappa_1,\kappa_2\cdots,\eta_{\alpha\beta},e_i^{\tck\alpha},
F_{ij}^{\tck\tck\alpha},F_{ij}^{\tck\tck\alpha\beta}\Big)\ .
\ee
Eq. (\ref{329}) shows that $({\cal H}_\alpha^{\tck ij},
{\cal H}_{\alpha\beta}^{\,\tck\tck ij})$ are both tensor densitites,
a fact which was not obvious in their definition (\ref{324}).

After (\ref{218}) we saw already that
$(F_{\alpha\beta}^{\tck\tck\gamma},F_{\alpha\beta}^
{\tck\tck\gamma\delta})$ are P-tensors. Consequently (\ref{330}) can
be simplified and the most general first order gauge field
lagrangian reads
\be
\label{331}
V=eV\Big(
\kappa_1,\kappa_2\cdots,\eta_{\alpha\beta},
F_{\alpha\beta}^{\tck\tck\gamma},F_{\alpha\beta}^
{\tck\tck\gamma\delta}\Big)\ .
\ee
It is remarkable that the potentials don't appear explicitly in $V$.

Let us now collect the coefficients of the $(D_i\varepsilon^\alpha,
D_i\omega^{\alpha\beta})$-terms in (\ref{325}). The calculation
yields
\be
\label{332}
\frac{\delta V}{\delta e_i^{\tck\alpha}}\equiv
-D_j{\cal H}_\alpha^{\tck ij}+\varepsilon_\alpha^{\tck i}
\ee
and
\be
\label{333}
\frac{\delta V}{\delta \Gamma_i^{\tck\alpha\beta}}\equiv
-D_j{\cal H}_{\alpha\beta}^{\,\tck\tck ij}+\varepsilon_
{\alpha\beta}^{\tck\tck i}
\ee
with
\be
\label{334}
\varepsilon_\alpha^{\tck i}:=\ \varepsilon_{\tck\alpha}^i
V-F_{\alpha j}^{\tck\tck\gamma}{\cal H}_\gamma^{\tck ji}
-F_{\alpha j}^{\tck\tck\gamma\delta}{\cal H}_{\gamma\delta}^{\tck\tck ji}
\ee
and
\be
\label{335}
\varepsilon_{\alpha\beta}^{ \tck\tck i}:=\
{\cal H}_{[\beta\alpha]}^{\ \tck\tck \ i}\ ,
\ee
respectively. Of course, the quantities (\ref{334}), (\ref{335}) are
well-behaved tensor densities.

The interpretation of (\ref{335}) is obvious. Because of (\ref{324})
we find
\be
\label{336}
\varepsilon_{\alpha\beta}^{ \tck\tck i}:=
e_{k[\alpha}{\cal H}_{\beta]}^{\tck \ ki}=
\frac{\partial V}{\partial\partial_i
e_k^{\tck[\beta}}e_{|k|\alpha]}=
\frac{\partial V}{\partial\partial_i
e_k^{\tck\gamma}}f_{\beta\alpha}e_k^{\tck\gamma}\ .
\ee
A comparison with (\ref{39}) shows that (\ref{336}) represents the
canonical spin current of the translational gauge potential
$e_i^{\tck\alpha}$. Analogously, (\ref{334}) has the structure and the
dimension of a canonical momentum current of both potentials
$(e_i^{\tck\alpha},\Gamma_i^{\tck\alpha\beta})$, as is evidenced by a
comparison with (\ref{38}).

We don't need the identities resulting from the
$(\varepsilon^\alpha,\omega^{\alpha\beta})$-terms, since they will
become trivial consequences of the field equations (\ref{344}),
(\ref{345}) as substituted into the conservation laws (\ref{310}),
(\ref{311}).

\subsection{Gauge Field Equations}

The total action function of the interacting matter and gauge fields reads
\bea
\label{337}
W&=&\int d^4x\Big[{\cal L}\Big(
\eta_{\alpha\beta},\gamma^\alpha\cdots,\psi,\partial_i\psi,e_i^{\tck\alpha},
\Gamma_i^{\tck\alpha\beta}\Big)+\nonumber\\
&&+V\Big(
\kappa_1,\kappa_2\cdots,\eta_{\alpha\beta},e_i^{\tck\alpha},
\Gamma_i^{\tck\alpha\beta},\partial_ke_i^{\tck\alpha},
\partial_k\Gamma_i^{\tck\alpha\beta}\Big)\Big]\ .
\eea
Local P-invariance implies two things, inter alia: It yields, as
applied to ${\cal L}$, the minimal coupling prescription (\ref{315}),
and it leads, as applied to $V$, to the gauge field lagrangian
(\ref{331}) as the most general one allowed. Consequently we have
\bea
\label{338}
W&=&\int d^4x\Big[L\Big(
\eta_{\alpha\beta},\gamma^\alpha\cdots,\psi,D_\alpha\psi
\Big)+\nonumber\\
&&+V\Big( \kappa_1,\kappa_2\cdots,\eta_{\alpha\beta},
F_{\alpha\beta}^{\tck\tck\gamma},
F_{\alpha\beta}^{\tck\tck\gamma\delta}\Big)\Big]\ .
\eea

The independent variables are\symbolfootnote[1] {Independent variation
of tetrad and connection is usually and mistakenly called ``Palatini
Variation". In order to give people who only cite Palatini's famous
paper a chance to really read it, an English translation is provided in
this volume on my suggestion. Sure enough, this service will not change
habits.}
\be
\label{339}
\psi=\textrm{matter field},\quad
(e_i^{\tck\alpha},\Gamma_i^{\tck\alpha\beta})=
\textrm{gauge potentials}.
\ee
The action principle requires
\be
\label{340}
\delta_{\psi,e,\Gamma}W=0\ .
\ee
We find successively
\be
\label{341}
\frac{\delta{\cal L}}{\delta\psi}=0\ ,
\ee
see (\ref{32}), and
\bea
\label{342}
&&-\frac{\delta V}{\delta e_i^{\tck\alpha}}=
\frac{\delta {\cal L}}{\delta e_i^{\tck\alpha}}\ , \\
\label{343}
&&-\frac{\delta V}{\delta \Gamma_i^{\tck\alpha\beta}}=
  \frac{\delta {\cal L}}{\delta \Gamma_i^{\tck\alpha\beta}}\ .
\eea
By (\ref{38}), (\ref{39}) the right hand sides of (\ref{342}),
(\ref{343}) are identified as material momentum and spin currents,
respectively. Their left hand sides can be rewritten using the
tensorial decompositions (\ref{332}), (\ref{333}). Therefore we get the
following two gauge field equations:
\bea
\label{344}
&&D_j{\cal H}_\alpha^{\tck ij}-\varepsilon_\alpha^{\tck i}=
  e\Sigma_\alpha^{\tck i}\ ,                    \\
\label{345}
&&D_j{\cal H}_{\alpha\beta}^{\,\tck\tck ij}
  -\varepsilon_{\alpha\beta}^{\tck\tck i}
  =e\tau_{\alpha\beta}^{\tck\tck i}\ .
\eea
We call these equations 1st (or translational) field equation and 2nd
(or rotational) field equation, respectively, and we remind ourselves
of the following formulae relevant for the field equations, see
(\ref{329}), (\ref{334}), (\ref{335}):
\bea
\label{346}
&&{\cal H}_\alpha^{\tck ij}=2
  \frac{\partial V}{\partial F_{ji}^{\tck\tck\alpha}};\quad
  {\cal H}_{\alpha\beta}^{\,\tck\tck ij}=2
  \frac{\partial V}{\partial F_{ji}^{\tck\tck\alpha\beta}}\ ,\\
\label{347}
&&\varepsilon_\alpha^{\tck i}=e_{\tck\alpha}^iV-
  F_{\alpha j}^{\tck\tck\gamma}{\cal H}_\gamma^{\tck ji}-
  F_{\alpha j}^{\tck\tck\gamma\delta}{\cal H}_{\gamma\delta}^
  {\tck\tck ji}\ ,                                         \\
\label{348}
&&\varepsilon_{\alpha\beta}^{\tck\tck i}=
  {\cal H}_{[\beta\alpha]}^{\ \tck\tck \ i}\ .
\eea

In general, without specifying a definite field lagrangian $V$, the
field momenta ${\cal H}_\alpha^{\tck ij}$ and ${\cal
H}_{\alpha\beta}^{\,\tck\tck ij}$ are of first order in their
corresponding potentials $e_i^{\tck\alpha}$ and
$\Gamma_i^{\tck\alpha\beta}$, i.e. (\ref{344}) and (\ref{345}) are
generally second order field equations for $e_i^{\tck\alpha}$ and
$\Gamma_i^{\tck\alpha\beta}$,
\be
\label{349}
\left\{
\begin{array}{rl}
\partial\partial e+\cdots \sim\Sigma,\\[4pt]
\partial\partial \Gamma+\cdots \sim\tau.
\end{array} \right\}
\ee
Furthermore the currents $\Sigma_\alpha^{\tck i}$ and
$\tau_{\alpha\beta}^{\tck\tck i}$ couple to the potentials
$e_i^{\tck\alpha}$ and $\Gamma_i^{\tck\alpha\beta}$, respectively.
Clearly then the field equations are of the Yang-Mills type (cf. eq.
(12a) in \cite{7}), as expected in faithfully executing the gauge idea.

There is the fundamental difference, however. The universality of the
P-group induces the existence of the tensorial currents
$(\varepsilon_\alpha^{\tck i},\varepsilon_{\alpha\beta}^ {\tck\tck i})$
of the gauge potentials themselves.\symbolfootnote[1] {As a by-product
of our investigations, we found the energy-momentum tensor
$\varepsilon_\alpha^{\tck i}$ of the gravitational field. Hence the
tetrad (or rather $T_4$-) people who searched for this quantity for
quite a long time, were not all that wrong, as will become clear from
the lecture of Nitsch \cite{16}. As we will see in Section 4.4, for a
$V$ linear in curvature, $\varepsilon_\alpha^{\tck i}$ turns out to be
just the Einstein tensor of the $U_4$, for a quadratic lagrangian
$\varepsilon_\alpha^{\tck i}$ is of the type of a contracted
Bel-Robinson tensor (see \cite{49}) or, to speak in electrodynamical
terms, of the type of Minkowski's energy-momentum tensor.} In other
words, in the PG it is not only the material currents
$(e\Sigma_\alpha^{\tck i},e\tau_{\alpha\beta}^{\tck\tck i})$ which
produce the fields, but rather the sum of the material and the gauge
currents $(e\Sigma_\alpha^{\tck i} +\varepsilon_\alpha^{\tck
i},e\tau_{\alpha\beta}^{\tck\tck i} +\varepsilon_{\alpha\beta}^
{\tck\tck i})$, this sum being a tensor density again.

Taking into account (\ref{336}) and the discussion following it, it is
obvious that $\varepsilon_\alpha^{\tck i}$ is the \emph{momentum
current} (energy-momentum density) and $\varepsilon_{\alpha\beta}^
{\tck\tck i}$ the \emph{spin current} (spin angular momentum density)
\emph{of the gauge fields}. Whereas both gauge potentials carry
momentum, as is evident from (\ref{347}), only the translational
potential $e_i^{\tck \alpha}$ gives rise to a tensorial spin, as one
would expect, according to (\ref{226}), (\ref{227}), from the behavior
of the $(e_i^{\tck\alpha},\Gamma_i^{\tck\alpha\beta})$-set under local
rotations.

Hence the rotational potential as a quasi-intrinsic gauge potential has
vanishing dynamical spin in the sense of the PG, and this fact goes
well together with the vanishing dynamical spin of the Maxwell field
$A_i$.

Within the objective of finding a genuine gauge field theory of the
P-group, the structure put forward so far, materializing in particular
in the two general field equations\symbolfootnote[2] {ln words we could
summarize the structure of the field equations as follows: gauge
covariant divergence of field momentum = (gauge + material) current. In
a pure Yang-Mills theory, just omit the phrase ``gauge +".}
(\ref{344}), (\ref{345}), seems to us final. The picking of a suitable
field lagrangian, which is the last step in establishing a physical
theory, is where the real disputation sets in (see Lecture 4).

To our knowledge there exist only the following objections
against the PG:
\begin{itemize}

\item[--] The description of matter in terms of classical fields is
illegitimate. This is a valid objection which can be met
by \emph{quantizing} the theory.

\item[--] The PG is too special, it needs \emph{grading}, i.e.
instead of a PG we should rather have a graded PG. Then we end up with
supergravity in a space with supertorsion and supercurvature (see
\cite{14,50,51}). Up to now, it is not clear whether this step is
really compulsory.

\item[--] The PG is too special, it needs the extension to the gauge
theory of the 4-dimensional real affine group GA(4,R)
(\emph{metric-affine} theory). There are in fact a couple of
independent indications pointing in this direction. Therefore we
developed a tensorial \cite{52} and a spinorial \cite{53,54,55} version
of such a theory. What is basically happening is that the 2nd field
equation (\ref{345}) loses its antisymmetry in $\alpha\beta$, i.e. new
intrinsic material currents (dilation plus shear) which, together with
spin, constitute the hypermomentum current, couple to the newly
emerging gauge potential $\Gamma_i^{\tck(\alpha\beta)}$.

\item[--] The PG is too special, it needs the extension to a GA(4,R)-
gauge \emph{and} it needs grading as well (\cite{56} and refs.
therein). May be.
\end{itemize}

Any of these objections, however, doesn't make a thorough
investigation into the PG futile, it is rather a prerequisite for
a better understanding of the extended frameworks.


\def\ver{\phantom{$\left\vert\vbox to 1.2em{}\right.\hspace{-.5em}$}}

\begin{landscape}
\begin{table}[htb]
\subsection{The Structure of Poincar\'e Gauge Field Theory Summarized}

\begin{tabular}{|l|l|l|l||l|}
\hline
& & translation\ver & rotation & phase change U(1)\\
\hline\hline

\multirow{4}{*}{\raisebox{-0em}[0pt]{Geometry}}  
& infinitesimal generator\ver& $D_\alpha \hspace{30pt} 4$
  & $f_{\alpha\beta} \hspace{40pt} 6$ & $q$\hspace{22pt}1 charge\\
                                                             \cline{2-5}
& gauge potential\ver & $e_i^{\tck\alpha}$ \hspace{13pt} tetrad
  & $\Gamma_i^{\tck\alpha\beta}$ \hspace{9pt} connection & $A_i$\\
                                                             \cline{2-5}
& gauge field strength\ver & $F_{ij}^{\tck\tck\alpha}$ ~~torsion
  & $F_{ij}^{\tck\tck\alpha\beta}$ ~~curvature & $F_{ij}$\\ \cline{2-5}
& Bianchi identity\ver
  & $D_{[i}F_{jk]}^{\tck\tck \ \alpha}\equiv
                                      F_{[ijk]}^{\tck\tck\tck \ \alpha}$
  & $D_{[i}F^{\tck\tck\,}_{jk]}{}^{\alpha\beta}\equiv 0$
  & $\partial_{[i}F_{jk]}\equiv0$\\
\hline\hline

\multirow{2}{*}{\raisebox{-0em}[0pt]{Kinematics}} 
& material current\ver
  & $\hat\Sigma_\alpha^{\tck i}=\delta{\cal L}/\delta e_i^{\tck\alpha}$
  & $\hat\tau_{\alpha\beta}^{\tck\tck i}=\delta{\cal L}/
        \delta \Gamma_i^{\tck\alpha\beta}$
  & ${\cal J}^i=\delta{\cal L}/\delta A_i$\\ \cline{2-5}
& conservation law\ver & $D_i\hat\Sigma_\alpha^{\tck i}=
      F_{\alpha i}^{\tck\tck\beta}\hat\Sigma_\beta^{\tck i}+
      F_{\alpha i}^{\tck\tck\beta\gamma}\hat\tau_{\beta\gamma}^{\tck\tck i}$
  & $D_i\hat\tau_{\alpha\beta}^{\tck\tck i}-\hat\Sigma_{[\alpha\beta]}=0$
  & $\partial_j{\cal J}^i=0$\\
\hline\hline

\multirow{2}{*}{\raisebox{-0em}[0pt]{Dynamics}}  
& field momentum\ver & ${\cal H}_\alpha^{\tck ij}=2\partial V/\partial
        F_{ji}^{\tck\tck\alpha}$
  & ${\cal H}_{\alpha\beta}^{\,\tck\tck ij}=2\partial V/\partial
        F_{ji}^{\tck\tck\alpha\beta}$
  & ${\cal H}^{ij} =2\partial V_{\textrm{Max}}/\partial F_{ji}$\\
                                                           \cline{2-5}
& field equation\ver & $D_j{\cal H}_\alpha^{\tck ij}
    -\varepsilon_\alpha^{\tck i}=\hat\Sigma_\alpha^{\tck i}$
  & $D_j{\cal H}_{\alpha\beta}^{\,\tck\tck ij}-
  \varepsilon_{\alpha\beta}^{\tck\tck i}=\hat\tau_{\alpha\beta}^{\tck\tck i}$
  & $\partial_j{\cal H}^{ij}={\cal J}^i$\\
& & first (translational)\ver & second (rotational) & Inhom. Maxwell\\
                                                           \cline{2-5}
& ECSK choice\ver & ${\cal H}_\alpha^{\tck ij}\equiv0$
  & ${\cal H}_{\alpha\beta}^{\,\tck\tck ij}=ee^i_{\tck[\alpha}
      e^i_{\tck\beta]}/\ell^2$ & for vacuum\\\cline{2-4}
& our choice\ver & ${\cal H}_\alpha^{\tck ij}=e\big(F^{ij}_{\tck\tck\alpha}
    +2e^{[i}_{\ \tck\alpha}F^{j]\gamma}_{\tck\ \tck\gamma}\big)/\ell^2$
  & ${\cal H}_{\alpha\beta}^{\,\tck\tck ij}
                            =eF^{ij}_{\tck\tck\alpha\beta}/\kappa$
  & ${\cal H}^{ij}=\hat\mu F^{ij}$\\
\hline
\end{tabular}
\caption{\rightskip 5pt
{\small (The caret ``\^{}" means that the quantity be
multiplied by $e=\textrm{det} \ e_i^{\tck\alpha}$. The momentum current
of the field is denoted by $\varepsilon_\alpha^{\tck i}$, see
(\ref{347}), the spin current by $\varepsilon_{\alpha\beta}^{\tck\tck
i}:\ ={\cal H}_{[\beta\alpha]}^ { \ \tck\tck \  i}$.) The
Riemann-Cartan geometry is dictated by a proper application of the
gauge idea to the Poincar\'e group. To require at that stage a
constraint like $F_{ij}^{\tck\tck\alpha\beta}=0$ (teleparallelism) or
$F_{ij}^{\tck\tck\alpha}=0$ (riemannian geometry) would seem without
foundation. Provided one takes a first-order lagrangian of the type
${\cal L}(\psi,\partial\psi,e,\Gamma) +V(e,\Gamma,\partial
e,\partial\Gamma)$, the coupling of matter to geometry and the two
gauge field equations are inequivocally fixed and we find
$V=eV(F_{\alpha\beta}^{\tck\tck\gamma},
F_{\alpha\beta}^{\tck\tck\gamma\delta})$.}
}\vspace{-2.5pt}
\end{table}
\end{landscape}

\lecture{Lecture 4: Picking a Gauge Field Lagrangian}
\addcontentsline{toc}{section}{Lecture 4:~ Picking a Gauge Field
Lagrangian}

Let us now try to find a suitable gauge field lagrangian in
order to make out of our PG-framework a realistic physical theory.

\subsection{Hypothesis of Quasi-Linearity}

The leading terms of our field equations are
\bea
\label{41}
&&\partial_i{\cal H}_\alpha^{\tck ij}\sim \Sigma_\alpha^{\tck i}\ ,\\
\label{42}
&&\partial_j{\cal H}_{\alpha\beta}^{\,\tck\tck ij}\sim \tau_{\alpha\beta}
^{\tck\tck i}\ .
\eea
In general the field momenta will depend on the same variables as $V$:
\bea
\label{43}
&&{\cal H}_\alpha^{\tck ij}={\cal H}_\alpha^{\tck ij}
(\kappa_1,\kappa_2\cdots,\eta,e,\Gamma,\partial e,\partial\Gamma)\,,\\[3pt]
\label{44}
&&{\cal H}_{\alpha\beta}^{\,\tck\tck ij}={\cal H}_{\alpha\beta}
^{\,\tck\tck ij}
(\kappa_1,\kappa_2\cdots,\eta,e,\Gamma,\partial e,\partial\Gamma)\ .
\eea
The translational momentum, being a third-rank tensor, cannot depend
on the derivatives of the connection $(\partial_\ell\Gamma_k
^{\tck\gamma\delta})$, because this expression is of even rank.
Furthermore, algebraic expressions of $\Gamma_k
^{\tck\gamma\delta}$ never make up a tensor. Hence we have
\be
\label{45}
{\cal H}_\alpha^{\tck ij}={\cal H}_\alpha^{\tck ij}
\Big(\kappa_1,\kappa_2\cdots,\eta_{\gamma\delta},e_k^{\tck\gamma},
F_{k\ell}^{\tck\tck\gamma}\Big)\ .
\ee
Similarly we find for the rotational momentum
\be
\label{46}
{\cal H}_{\alpha\beta}^{\,\tck\tck ij}={\cal H}_{\alpha\beta}
^{\,\tck\tck ij}
\Big(\kappa_1,\kappa_2\cdots,
\eta_{\gamma\delta},e_k^{\tck\gamma},
F_{k\ell}^{\tck\tck\gamma\delta}\,,
\Big[F_{k\ell}^{\tck\tck\gamma}\Big]^2\Big)\ .
\ee
This time, both curvature and torsion are allowed, the torsion
must appear at least as a square, however.

In order to narrow down the possible choices of a gauge field
lagrangian, we will assume quasi-linearity of the field equations
as a working hypothesis. This means that the second derivatives
in (\ref{41}), (\ref{42}) must only occur linearly. To our knowledge,
any successful field theory developed so far in physics obeys this
principle.\symbolfootnote[1]
{Instead of the quasi-linearity hypothesis one would prefer having
theorems of the Lovelock type available, see Aldersley \cite{57}
and references given there.} As a consequence the derivatives of
$(e_k^{\tck\gamma}, \Gamma_k^{\tck\gamma\delta})$ in (\ref{45}),
(\ref{46}) can only occur linearly or, in other words, (\ref{45}) and
(\ref{46}) are linear in torsion and curvatures, respectively. For
the translational momentum we have
\be
\label{47}
{\cal H}_\alpha^{\tck ij}\sim g \Big(\kappa_1,\kappa_2\cdots,
\eta_{\gamma\delta},e_k^{\tck\gamma}\Big)+
\textrm{lin}_t\Big(\kappa_1,\kappa_2\cdots,
\eta_{\gamma\delta},e_k^{\tck\gamma},
F_{k\ell}^{\tck\tck\gamma}\Big)\ .
\ee
Out of $\eta_{\gamma\delta}$ and $e_k^{\tck\gamma}$ we cannot
construct a third rank tensor i.e. $g$ has to vanish. Accordingly we
find
\bea
\label{48}
&&{\cal H}_\alpha^{\tck ij}\sim
\textrm{lin}_t\Big(\kappa_1,\kappa_2\cdots,
\eta_{\gamma\delta},e_k^{\tck\gamma},
F_{k\ell}^{\tck\tck\gamma}\Big)\ ,              \\
\label{49}
&&{\cal H}_{\alpha\beta}^{\,\tck\tck ij}\sim h
\Big(\kappa_1,\kappa_2\cdots,
\eta_{\gamma\delta},e_k^{\tck\gamma}\Big)+
\textrm{lin}_c\Big(\kappa_1,\kappa_2\cdots,
\eta_{\gamma\delta},e_k^{\tck\gamma},
F_{k\ell}^{\tck\tck\gamma\delta}\Big)\,,\qquad
\eea
where lin$_t$ and lin$_c$ denote tensor densities being linear and
homogeneous in torsion and curvature, respectively. The possibility
of having the curvature-independent term $h$ in (\ref{49}) is again
a feature particular to the PG.

We note that the hypothesis of quasi-linearity constrains the
choice of the ``constitutive laws" (\ref{48}), (\ref{49}) and
of the corresponding gauge field lagrangian appreciably. The lagrangian
$V$, as a result of (\ref{346}) and of (\ref{48}), (\ref{49}), is
at most quadratic in torsion and curvature (compare (\ref{331})):
\be
\label{410}
V\sim e(\textrm{const} + \textrm{torsion}^2 + \textrm{curvature}
+ \textrm{curvature}^2)\ .
\ee
By the definition of the momenta, we recognize the following
correspondence between (\ref{410}) and (\ref{48}), (\ref{49}):
\bea
\label{411}
&&\textrm{const} \to {\cal H}_\alpha^{\tck ij}\equiv 0\qquad
{\cal H}_{\alpha\beta}^{\tck\tck ij}\equiv 0\ ,  \\
\label{412}
&&\textrm{torsion}^2\to \textrm{lin}_t\ ,        \\
\label{413}
&&\textrm{curvature}\to h\ ,                     \\
\label{414}
&&\textrm{curvature}^2\to \textrm{lin}_c\ .
\eea

The correspondence (\ref{411}) can be easily understood. For
vanishing field momenta we find (-const)$e^i_{\tck\alpha}=
\Sigma_\alpha^{\tck i}$ and $\tau_{\alpha\beta}^{\tck\tck i}=0$.
Clearly then this term in $V$ is of the type of a
cosmological-constant-term in GR, i.e. $V=e\times$const. doesn't
make up an own theory, it can only supplement another lagrangian.

In regard to (\ref{413}) we remind ourselves that in a $U_4$ we
have the identities
\be
\label{415}
F_{ij}^{\tck\tck\alpha\beta}\equiv
F_{[ij]}^{\tck\tck\,\alpha\beta}\equiv
F_{ij}^{\tck\tck[\alpha\beta]}\ .
\ee
Hence, apart from a sign difference, there is only one way to
contract the curvature tensor to a scalar:
\be
\label{416}
F:= e^i_{\tck\beta}e^j_{\tck\alpha}F_{ij}^{\tck\tck\alpha\beta}
= e^i_{\tck[\beta}e^j_{\tck\alpha]}F_{ij}^{\tck\tck\alpha\beta}\ .
\ee
Hence the term linear in curvature in (\ref{410}) is just
proportional to the curvature scalar $F$.

Since we chose units such that $\hbar=c=1$, the dimension of
$V$ has to be (length)$^{-4}$:
\be
\label{417}
[V]=\ell^{-4}\ .
\ee
Furthermore we have
\be
\label{418}
\Big[e_i^{\tck\alpha}\Big]=1, \quad
\Big[\Gamma_i^{\tck\alpha\beta}\Big]=\ell^{-1}\ ,
\ee
and
\be
\label{419}
\Big[F_{ij}^{\tck\tck\alpha}\Big]=\ell^{-1}, \quad
\Big[F_{ij}^{\tck\tck\alpha\beta}\Big]=\ell^{-2}\ .
\ee
Accordingly a more definitive form of (\ref{410}) reads
\be
\label{420}
V\sim e\Big[\frac{1}{{\mathop{L}\limits_{0}}^4}+
\frac{1}{{\mathop{L}\limits_{1}}^2}(\textrm{torsion})^2+
\frac{1}{{\mathop{L}\limits_{2}}^2}(\textrm{curv.scalar})+
\frac{1}{\kappa}(\textrm{curvature})^2 \Big]
\ee
with $\mathop{[L]}\limits_{0}=\mathop{[L]}\limits_{1}=
\mathop{[L]}\limits_{2}=\ell,~ [\kappa]=1$. Of course, any number of
additi\-o\-nal dimensionless coupling constants are allowed in (\ref{420}).
If we put ~$\mathop{L}\limits_{0}=\mu L,~ \mathop{L}\limits_{1}=L,~
\mathop{L}\limits_{2}=\chi^{1/2}L$,~ then (\ref{420}) gets slightly
rewritten and we have
\be
\label{421}
V\sim e\Big[
\frac{1}{(\mu L)^4}+\frac{1}{L^2}\Big\{
(\textrm{torsion})^2+\frac{1}{\chi}
(\textrm{curv.scalar})\Big\}+\frac{1}{\kappa}
(\textrm{curvature})^2\Big]
\ee
with
\be
\label{422}
[L]=\ell,\quad [\mu]=[\chi]=[\kappa]=1\ .
\ee
Therefore (\ref{48}), (\ref{49}) finally read
\bea
\label{423}
&&e^{-1}{\cal H}_\alpha^{\tck ij}\sim
L^{-2}\textrm{lin}_t
\Big(d_1,d_2\cdots,\eta_{\gamma\delta},
e_k^{\tck\gamma},F_{k\ell}^{\tck\tck\gamma}\Big)\ , \\
\label{424}
&&e^{-1}{\cal H}_{\alpha\beta}^{\,\tck\tck\, ij}\sim
\chi^{-1}L^{-2}e^i_{\tck[\alpha}e^j_{\tck\beta]}
+\kappa^{-1}\textrm{lin}_c
\Big(f_1,f_2\cdots,\eta_{\gamma\delta},
e_k^{\tck\gamma},F_{k\ell}^{\tck\tck\gamma\delta}\Big),\qquad
\eea
with
\be
\label{425}
[d_1]=[d_2]=\cdots=[f_1]=[f_2]=\cdots=1\ .
\ee

Provided we don't only keep the curvature-square piece in the
lagrangian (\ref{421}) alone, which leads to a non-viable
theory,\symbolfootnote[1]
{$\dots$ to the Stephenson-Kilmister-Yang ansatz, see \cite{58}
and references given there.} we need one \emph{fundamental length},
three primary dimensionless constants $\mu$ (scaling the cosmological
constant), $\chi$ (fixing the relative weight between
torsion-square and curvature scalar), $\kappa$ (a measure of the
``roton" coupling), and a number of secondary dimensionless constants
$d_1,d_1\cdots$ and $f_1,f_2\cdots$.

We have discussed in this section, how powerful the quasi-linearity
hypothesis really is. It doesn't leave too much of a choice for the
gauge field lagrangian.

\subsection{Gravitons and Rotons}

We shall take the quasi-linearity for granted. Then, according to
(\ref{423}), (\ref{424}), the leading derivatives of the field
momenta in general are
\bea
\label{426}
&&{\cal H}_\alpha^{\tck ij}\sim L^{-2}\partial e\ , \\
\label{427}
&&{\cal H}_{\alpha\beta}^{\,\tck\tck ij}\sim \kappa^{-1}\partial\Gamma\ .
\eea
Substitute (\ref{426}), (\ref{427}) into the field equations
(\ref{344}), (\ref{345}) and get the scheme:
\be
\left.
\begin{array}{rl}
\partial\partial e+\cdots\sim L^2\Sigma,\\[4pt]
\partial\partial\Gamma+\cdots\sim \kappa\tau\ .
\end{array} \right\}
\ee
Let us just for visualization tentatively take the simplest
\emph{toy theory} possible for describing such a behavior, patterned
after Maxwell's theory,
\bea
\label{429}
&&{\cal H}_\alpha^{\tck ij}\sim L^{-2}F^{ij}_{\tck\tck\alpha}\ ,\\
\label{430}
&&{\cal H}_{\alpha\beta}^{\,\tck\tck ij}\sim\kappa^{-1}
F^{ij}_{\tck\tck\alpha\beta}\ ,
\eea
i.e.
\be
\label{431}
V\sim L^{-2}F^{ij}_{\tck\tck\alpha}F_{ij}^{\tck\tck\alpha}
+\kappa^{-1}
F^{ij}_{\tck\tck\alpha\beta}F_{ij}^{\tck\tck\alpha\beta}\ .
\ee
Then we have for both potentials ``kinetic energy" - terms in
(\ref{431}).
Observe that the index positions in (\ref{429}), (\ref{430})
are chosen in such a way that in (\ref{431}) only pure
squares appear of each torsion or curvature component,
respectively. This is really the simplest choice.

Clearly then, the PG-framework in its general form allows for two types
of propagating gauge bosons: gravitons $e^{\tck\alpha}_i$ (``weak
Einstein gravity") and rotons $\Gamma^{\tck\alpha\beta}_i$ (``strong
Yang-Mills gravity").\symbolfootnote[1] {The f-g-theory of gravity of
Zumino, Isham, Salam, and Strathdee (for the references see \cite{58})
appears more fabricated as compared to the PG. The term ``strong
gravity" we borrowed from these authors. There should be no danger that
our rotons be mixed up with those of liquid helium. Previously we
called them ``tordions" \cite{36,1}, see also Hamamoto \cite{59}, but
this gives the wrong impression as if the rotons were directly related
to torsion. With the translation potential $e^{\tck\alpha}_i$ there is
associated a set of 4 vector-bosons of dynamical spin 1. It would be
most appropriate to call these quanta ``translatons" since the graviton
is really a spin-2 object.} These and only these two types of
interactions are allowed and emerge quite naturally from our
phenomenological analysis of the P-group. \emph{We postulate that both
types of P-gauge bosons exist in nature} \cite{58}.

From our experience with GR we know that the fundamental length
$L$ of (\ref{422}) in the PG has to be identified
with the Planck length $\ell$ ($K$ = relativistic
gravitational constant)
\be
\label{432}
L=\ell=(K)^{1/2}\approx 10^{-32} \, \textrm{cm}\ ,
\ee
whereas we have no information so far on the magnitude of the
dimensionless constant $\kappa$ coupling the rotons to material spin.

Gravitons, as we are assured by GR exist, but the rotons need
experimental verification. As gauge particles of the Lorentz
(rotation-) group $SO(3,1)$, they have much in common with, say,
$SU(2)$-gauge bosons. They can be understood as arising from a
quasi-internal symmetry $SO(3,1)$. It is tempting then to relate
the rotons to strong interaction properties of matter. However,
it is not clear up to now, how one could manage to exempt the
leptons from roton interactions. One should also keep in mind
that the rotons, because of the close link between
$e^{\tck\alpha}_i$ and $\Gamma^{\tck\alpha\beta}_i$, have
specific properties not shared by $SU(2)$-gauge bosons, their
propagation equation, the second field equation, carries an
effective mass-term $\sim\kappa\ell^{-2}\Gamma$, for
example.\symbolfootnote[1]
{Some attempts to relate torsion to weak interaction, were
recently criticized by DeSabbata and Gasperini \cite{60}.}

Before studying the roton properties in detail, one has to come
up with a definitive field lagrangian.

\subsection{Suppression of Rotons I: Teleparallelism}

Since the rotons haven't been observed so far, one could try
to suppress them. Let us look how such a mechanism works.

By inspection of (\ref{41}), (\ref{42}) and of (\ref{423}),
(\ref{424}) we recognize that the second derivatives of
$\Gamma_k^{\tck\gamma\delta}$ enter the 2nd field equation by
means of the rotational momentum (\ref{424}), or rather by means
of the lin$_c$-term of it. Hence there exist two possibilities of
getting rid of the rotons: drop ${\cal H}_{\alpha\beta}^{\tck\tck ij}$
altogether or drop only its lin$_c$-piece. We'll explore the first
possibility in this section, the second one in Section 4.4.

Let us put
\be
\label{433}
{\cal H}_{\alpha\beta}^{\,\tck\tck ij}\equiv 0\ .
\ee
Then the field equations (\ref{344}), (\ref{345}) read\par
\vbox{
\begin{align*}\hspace{36pt}
     \left\{
\begin{array}{r@{}l}
  D_j{\cal H}_\alpha^{\tck\, ij}-e^i_{\tck\,\alpha}V
  +F_{\alpha j}^{\,\tck\tck\gamma}
  {\cal H}_\gamma^{\tck\, ji}&=e\Sigma_\alpha^{\tck\, i}\ ,\\[5pt]
{\cal H}_{[\alpha\beta]}^{\,\,\tck\,\tck\,\,i}&=
   e\tau_{\alpha\beta}^{\tck\tck\, i}\,.
\end{array}
     \right\} \quad\text{(inconsistent)}\hspace{58pt}  
\end{align*}
\vspace{-65pt}
\begin{eqnarray}
\label{434}
   &&\hfill  \\[5pt]
\label{435}
   && \hfill
\end{eqnarray}
}
\par\noindent
For vanishing material spin $\tau_{\alpha\beta}^{\tck\tck i}=0$,
however, the tetrad spin ~${\cal H}_{\alpha\beta}^{\,\tck\tck i}$~
would vanish, too, and therefore force the tetrads out of
business. We were left with the term
$-e^i_{\tck\alpha}V=e\Sigma_\alpha^{\tck i}$ related to the
cosmological constant, see (\ref{411}). Hence this recipe is not
successful.

But it is obvious that eq. \ref{434} is of the desired type,
becouse it is a second-order field equation in $e_k^{\tck\gamma}$.
We know from (\ref{423}) that ${\cal H}_\alpha^{\tck ij}$ can
only be linear and homogeneous in $F_{k\ell}^{\tck\tck\gamma}$.
Additionally one can accommodate the Planck length in the
constitutive law (\ref{423}). Consequently, the most general linear
relation
\be
\label{436}
\mathop{\cal H}\limits^{T}{}_\alpha^{\tck ij}=
e\Big(d_1F_{\tck\tck\alpha}^{ji}+d_2F_\alpha^{\tck[ij]}+
d_3e^{[i}_{\ \tck\alpha}F^{j]\gamma}_{\,\tck \, \tck\gamma}
\Big)/\ell^2\ ,
\ee
as substituted in (\ref{434}), would be just an einsteinian type of
field equation provided the constants $d_1, d_2$, and $d_3$ were
appropriately fixed.

Our goal was to suppress the rotons. In (\ref{256}) and
(\ref{257}) we saw that we can get rid of an independent connection in
a $T_4$ as well as in a $V_4$. In view of (\ref{436}) the choice of a
$V_4$ would kill the whole lagrangian. Therefore we have to turn to
a $T_4$ and we postulate the lagrangian
\be
\label{437}
\mathop{V}\limits^{T}=e
\Big(
d_1F_{ijk}F^{ijk}+d_2F_{kji}F^{ijk}+d_3F_{ik}^{\tck\tck k}
F_{\tck\tck\ell}^{i\ell}+
\Lambda_{\alpha\beta}^{\tck\tck ij}F_{ij}^{\tck\tck\alpha\beta}
\Big)/4\ell^2\ ,
\ee
with $\Lambda_{\alpha\beta}^{\tck\tck ij}$ as a lagrangian multiplier,
i.e. we have imposed onto (\ref{436}) and onto our $U_4$-spacetime
the additional requirement of vanishing curvature. The translational
momentum $\mathop{\cal H}\limits^{T}{}_\alpha^{\tck ij}$ is still
given by (\ref{436}), but the rotational momentum, against our
original intention, surfaces again:
\be
\label{438}
{\cal H}_{\alpha\beta}^{\,\tck\tck ij}=e\Lambda_{\alpha\beta}
^{\,\tck\tck ij}/4\ell^2\ .
\ee
To insist on a vanishing ${\cal H}_{\alpha\beta}^{\,\tck\tck ij}$ turns
out to be not possible in the end, but the result of our insistence is
the interesting teleparallelism lagrangian (\ref{437}).

The field equations of (\ref{437}) read (see \cite{35})
\bea
\label{439}
&&D_j\mathop{\cal H}\limits^{T}{}_\alpha^{\tck ij}-e_{\tck\alpha}^i
\mathop{V}\limits^{T}+F_{\alpha j}^{\tck\tck\gamma}
\mathop{\cal H}\limits^{T}{}_\gamma^{\ ij}=e\Sigma_\alpha^{\tck i}\ ,\\
\label{440}
&&D_j\Big(e\Lambda_{\alpha\beta}
^{\,\tck\tck ij}/2\ell^2\Big)-
\mathop{\cal H}\limits^{T}{}_{[\beta\alpha]}^{\ \tck\tck \ i}
=e\tau_{\alpha\beta}^{\tck\tck i}\ ,           \\
\label{441}
&&F_{ij}^{\tck\tck\alpha\beta}=0\ .
\eea
One should compare these equations with the inconsistent set
(\ref{434}), (\ref{435}). Observe that in (\ref{439}) the term in
$\mathop{V}\limits^{T}$ carrying the lagrangian multiplier vanishes
on substituting (\ref{441}), i.e. from the point of view of the 1st
field equation, its value is irrelevant.

We imposed a $T_4$-constraint onto spacetime by the lagrangian
multiplier term in (\ref{437}). In a $T_4$ the
$\Gamma_i^{\tck\alpha\beta}$ are made trivial. Therefore, for
consistency, we cannot allow spinning matter (other than as
test particles) in such a $T_4$; spin is coupled to the
$\Gamma_i^{\tck\alpha\beta}$, after all. Accordingly, we take
\emph{macroscopic} matter with vanishing spin
$\tau_{\alpha\beta}^{\tck\tck i}\equiv0$. Then (\ref{440}) is of no
further interest, the multiplier just balances the tetrad spin.
We end up with the ``tetrad field equation" in a teleparallelism
spacetime $T_4$
\bea
\label{442}
&&D_j\mathop{\cal H}\limits^{T}{}_\alpha^{\tck ij}-\frac{1}{4}
e_{\tck\alpha}^i\Big(
F_{k\ell}^{\tck\tck\gamma}\mathop{\cal H}\limits^{T}{}_\gamma
^{\tck\ell k}\Big)+F_{\alpha j}^{\tck\tck\gamma}
\mathop{\cal H}\limits^{T}{}_\gamma
^{\tck ji}=e\sigma_\alpha^{\tck i}\ ,  \\
\label{443}
&&F_{ij}^{\tck\tck\alpha\beta}=0\ ,
\eea
with $\mathop{\cal H}\limits^{T}{}_\alpha^{\tck ij}$ as given by
(\ref{436}).

As shown in teleparallelism theory,\symbolfootnote[1]
{Nitsch \cite{16} and references given there, compare also
Hayashi and Shirafuji \cite{61}, Liebscher \cite{62},
Meyer \cite{35}, M{\o}ller \cite{63} and Nitsch and Hehl \cite{64}.}
there exists a one-parameter
family of teleparallelism lagrangians all leading to the
Schwarzschild solution including the Birkhoff theorem and all in
coincidence with GR up to 4th post-newtonian order:
\be
\label{444}
d_1=-\frac{1}{2}(\lambda+1),\quad d_2=\lambda-1,\quad d_3=2\ .
\ee
We saw already in Section 3.3 that the equations of motion for
macroscopic matter in a $T_4$ coincide with those in GR. For all
practical purposes this whole class of teleparallelism theories
(\ref{444}) is indistinguishable from GR. The choice
$\lambda=0$ leads to a \emph{locally} rotation-invariant
theory which is exactly equivalent to GR.

Let us sum up: In suppressing rotons we found a class of
viable teleparallelism theories for macroscopic gravity (\ref{442}),
(\ref{443}) with (\ref{436}), (\ref{444}). According to
(\ref{437}), they derive from a torsion-square lagrangian
supplemented by a multiplier term in
order to enforce a $T_4$-spacetime. The condition $\lambda=0$ yields a
theory indistinguishable from GR.

\subsection{Suppression of Rotons II: The ECSK-Choice}

This route is somewhat smoother and instead of finding a $T_4$, we
are finally led to a $V_4$. As we have seen, there exists the option
of only dropping the curvature piece of (\ref{424}). This leads to
the ansatz (we put $\chi=1$):
\be
\label{445}
\hat{\cal H}_{\alpha\beta}^{\,\tck\tck ij}=ee^i_{\tck[\alpha}
e^j_{\tck\beta]}/\ell^2\ .
\ee
A look at the field equations (\ref{344}), (\ref{345}) convinces us
that we are not in need of a non-vanishing translational momentum
now:
\be
\label{446}
\hat{\cal H}_\alpha^{\tck ij}\equiv 0\ .
\ee
With (\ref{446}) the field equations reduce to
\bea
\label{447}
&&F_{\alpha j}^{\tck\tck\gamma\delta}
  \hat{\cal H}_{\gamma\delta}^{\tck\tck ji}
  -e_{\tck\alpha}^i\hat V=e\Sigma_\alpha^{\tck i}\ , \\[3pt]
\label{448}
&&D_j\hat{\cal H}_{\alpha\beta}^{\,\tck\tck ij}=
e\tau_{\alpha\beta}^{\tck\tck i}\ .
\eea
Substitution of (\ref{445}) and using (\ref{243}) yields
\def\bea{\begin{eqnarray} }           \def\eea{\end{eqnarray} }
\bea
&& F_{\gamma\alpha}^{\tck\tck i\gamma}-\frac{1}{2}e_{\tck\alpha}^i
F_{\gamma\delta}^{\tck\tck \delta\gamma}=\ell^2
\Sigma_\alpha^{\tck i}\ ,                         \label{449}\\
&&\frac{1}{2}F_{\alpha\beta}^{\tck\tck i}+e^i_{\tck[\alpha}
F_{\beta]\gamma}^{\,\tck \, \tck\gamma}=
\ell^2\tau_{\alpha\beta}^{\tck\tck i}\ .          \label{450}
\eea
These are the field equations of the Einstein-Cartan-Sciama-
Kibble (ECSK)-theory of gravity\symbolfootnote[1]
{Sciama, who was the first to derive the field equations (\ref{449}),
(\ref{450}), judges this theory from today's point of view as follows
(private communication): ``The idea that spin gives rise to torsion
should not be regarded as an ad hoc modification of general relativity.
On the contrary, it has a deep group theoretical and
geometric basis. If history had been reversed and the spin of the
electron discovered before 1915, I have little doubt that Einstein
would have wanted to include torsion in his original formulation
of general relativity. On the other hand, the numerical differences
which arise are normally very small, so that the advantages
of including torsion are entirely theoretical."} derivable from the
lagrangian
\be
\label{451}
\hat V=\frac{1}{2}F_{ji}^{\tck\tck\alpha\beta}
\hat{\cal H}_{\alpha\beta}^{\,\tck\tck ij}=
ee^i_{\tck[\alpha}
e^j_{\tck\beta]}F_{ji}^{\tck\tck\alpha\beta}/2\ell^2=
F/2\ell^2\ .
\ee
The ECSK-theory has sa small additional contact interaction as
compared to GR.\symbolfootnote[2]
{A certain correspondence between the ECSK-theory and GR was
beautifully worked out by Nester \cite{65}. For a recent
analysis of the ECSK-theory see Stelle and West \cite{66,67}.}
For vanishing matter spin we recover GR. Hence
in this framework we got rid of an independent
$\Gamma_i^{\tck\alpha\beta}$ in a $V_4$ spacetime
in consistency with (\ref{257}).

Observe that something strange happened in (\ref{449}), (\ref{450}):
The rotation field strength $F_{ij}^{\tck\tck\alpha\beta}$ is
controlled by the translation current (momentum), the translation field
strength $F_{ij}^{\tck\tck\alpha}$ by the rotation current (spin). It
is like putting ``Chang's cap on Li's head" \cite{23}.

Linked with this intertwining of translation and rotation is
a fact which originally led Cartan to consider such a type of theory:
In the ECSK-theory the \emph{contracted Bianchi} identities
(\ref{215}), (\ref{216})
are, upon substitution of the field equations (\ref{449}),
(\ref{450}), identical to the conservation laws (\ref{310}),
(\ref{311}), for details see \cite{1}. From a gauge-theoretical
point of view this is a pure coincidence.
Probably this fact is a distinguishing feature of the ECSK-theory
as compared to other theories in the PG-framework.

Consequently a second and perhaps more satisfactory procedure
for suppressing rotons consists in picking a field lagrangian
proportional to the $U_4$-curvature scalar.

\subsection{Propagating Gravitons and Rotons}

After so much suppression it is time to liberate the rotons.
How could we achieve this goal? By just giving them enough kinetic
energy $\sim[(\partial e)^2 + (\partial\Gamma)^2]$ in order to
enable them to get away.

Let us take recourse to our toy theory (\ref{429}), (\ref{430}),
(\ref{431}). The lagrangian (\ref{431}) carries kinetic energy of both
potentials, and the gravitational constant, or rather the Planck
length, appears, too. But the game with the teleparallelism theories
made us wiser. The first term on the right hand side of (\ref{431})
would be inconsistent with macroscopic gravity, as can be seen from
(\ref{437}) with (\ref{444}). We know nothing about the
curvature-square term, hence we don't touch it and stick with the
simplest choice. Consequently the ansatz
\bea
\label{452}
&&{\cal H}_\alpha^{\tck ij}=e\Big[-\frac{1}{2}(\lambda+1)
F_{\tck\tck\alpha}^{ji}+(\lambda-1)F_\alpha^{\tck[ij]}
+2e^{[i}_{\ \tck\alpha}F^{j]\gamma}_{ \ \tck \ \tck\gamma}\Big] \\
&&\hspace{2.1em}=\langle\, e\ell^{-2}\eta_{\alpha\gamma}g^{k[i}g^{j]\ell}\,
  \rangle\Big[ -\frac{1}{2}(\lambda+1)F_{\ell k}^{\tck\tck\gamma}
  +(\lambda-1)F^\gamma_{\tck k\ell}
  +2e_k^{\tck\gamma}F_{\ell\delta}^{\tck\tck\delta}\Big]\ ,\nonumber\\
\label{453}
&&{\cal H}_{\alpha\beta}^{\,\tck\tck ij}
  =eF_{\tck\tck\alpha\beta}^{ij}/\kappa\ ,
\eea
or, by Euler's theorem for homogeneous functions, the corresponding
field lagrangian,
\bea
\label{454}
V &=& \frac{1}{4}\Big[F_{ji}^{\tck\tck\alpha}{\cal H}_{\alpha}^{\tck ij}
+F_{ji}^{\tck\tck\alpha\beta}{\cal H}_{\alpha\beta}^{\,\tck\tck ij}
\Big]\nonumber \\
&=&(e/4\ell^2)\Big[-\frac{1}{2}(\lambda+1)
F^{\tck\tck\alpha}_{ij}F_{\tck\tck\alpha}^{ij}+
(\lambda-1)
F^{\tck\tck\alpha}_{ij}F_{\alpha}^{\tck ji}+
2F_{\tck\tck\beta}^{i\beta}F^{\tck\tck\gamma}_{i\gamma}\Big]+\nonumber\\
&&+(e/4\kappa)\Big[
-F_{ij}^{\tck\tck\alpha\beta}F^{ij}_{\tck\tck\alpha\beta}\Big]\ ,
\eea
would appear to be simplest choice which encompasses macroscopic gravity
in some limit.

Consider ``constitutive assumption" (\ref{452}). In analogy with
Maxwell's theory one would like to have the metric appearing only
in the ``translational permeability" $\langle\rangle$ and not in the
$(\lambda-1)$-term in the bracket:
$F^\gamma_{\tck k\ell}=g_{\ell m}g^{no}e^m_{\tck\delta}e_n^{\tck\gamma}
F_{ok}^{\tck\tck\delta}$. For harmony one would then like to cancel this
term by putting $\lambda=1$. This is \emph{our choice}. Substitute
$\lambda=1$ into (\ref{452}) and use (\ref{243}). Then our
translational momentum can be put into a very neat form:
\be
\label{455}
{\cal H}_\alpha^{\tck ij}=e\Big(F_{\tck\tck\alpha}^{ij}+2e^{[i}_
{\,\tck\alpha}F^{j]\gamma}_{\,\tck \,\tck\gamma}\Big)/\ell^2=
2e^{i\mu}e^{j\nu}e_{m\alpha}D_n\Big(
ee^m_{\tck[\mu}e^n_{\tck\nu]}\Big)/\ell^2\ .
\ee
There is another choice which is distinguished by some property.
This is the choice $\grave{\textrm{a}}$ la Einstein $\lambda=0$, since
$\lambda=0$ leads, if one enforces a $T_4$, to a locally
rotation-invariant theory, as was remarked on in
Section 4.3.\symbolfootnote[1]
{After the proposal \cite{24,58} of the lagrangian with $\lambda=1$,
Rumpf \cite{68} worked out an analysis of the lagrangian in terms
of differential forms and formulated a set of guiding principles
yielding the $(\lambda =1)$-choice. It was also clear from his work
that this choice is the most natural one obeying (\ref{444})
from a gauge-theoretical point of view. Recently Wallner \cite{69},
in a most interesting paper, advocated the use of $\lambda = 0$.}
Apart from these two possibilities,
there doesn't exist to our knowledge any other preferred choice
of $\lambda$.

Before we substantiate our choice by some deeper-lying ideas,
let us look back to our streamlined toy theory (\ref{454}) cum
(\ref{452}), (\ref{453}) and compare it with the general
quasi-linear structure (\ref{421}) cum (\ref{423}), (\ref{424}):

Since we want propagating rotons, the curvature-square term in
(\ref{421}) is indispensable, i.e. $\kappa$ is finite.
We could accommodate more dimensionless constants by looking for
the most general linear expression lin$_c$ in (\ref{424}).
Furthermore, if we neglect the cosmological
term, then there is only to decide of how to put gravitons
into the theory, either by means of the torsion-square term or by
means of the curvature scalar (or with both together). Now, curvature
is already taken by the roton interaction in the quadratic
curvature-term, i.e. the rotons should be suppressed in the limit
of vanishing curvature (then the rotons' kinetic energy is zero).
In other words, curvature is no longer at our disposal and the
torsion-square piece has to play the role of the gravitons' kinetic
energy. And we know from teleparallelism that it can do so. Since
we don't need the curvature scalar any longer, we drop it and put
$\chi=\infty$, even if that is not necessarily implied by the arguments
given. It seems consistent with this picture that theories in a
$U_4$ with (curvature scalar) + (curvature)$^2$ don't seem to have a
decent newtonian limit.\symbolfootnote[2]
{Theories of this type have been investigated by Anandan \cite{70,71},
Fairchild \cite{72}, Mansouri and Chang \cite{73}, Neville \cite{74,75},
Ramaswamy and Yasskin \cite{76}, Tunyak \cite{21} and others.}

Collecting all these arguments, we see that the gauge field
lagrangian (\ref{454}) has a very plausible structure both from the
point of view of allowing rotons to propagate and of being
consistent with macroscopic gravity in an enforced $T_4$-limit.

\subsection{The Gordon Decomposition Argument}

The strongest argument in favor of the choice with
$\lambda=1$ \cite{24,58,13,25} comes from other quarters, however.
Take the lagrangian ${\cal L}$ of a Dirac field
$\psi(x)$ and couple it minimally to a $U_4$ according
to the prescription (\ref{315}). Compute the generalized Dirac equation
according to (\ref{32}) and the momentum and spin currents according to
(\ref{38}) and (\ref{39}), respectively \cite{77}. We find
\bea
\label{456}
&&\Sigma_\alpha^{\tck i}=\frac{i}{2}\bar\Psi\gamma^iD_\alpha\psi+h.c.\,\\
\label{457}
&&\tau_{\alpha\beta}^{\tck\tck i}=\frac{i}{2}\bar\Psi\gamma^i
f_{\alpha\beta}\psi+h.c.\ .
\eea
($i$= imaginary unit, $h.c.$= hermitian conjugate, $\bar\psi$= Dirac
adjoint).

Execute a Gordon decomposition of both currents and find
(we will give here the results for an $M_4$, they can be
readily generalized to a $U_4$):
\bea
\label{458}
&&\Sigma_\alpha^{\tck i}=\mathop{\Sigma_\alpha^{\tck i}}
\limits^{\textrm{conv}}+\partial_j\Big(
M^{ij}_{\tck\tck\alpha}+2
e^{[i}_{\,\tck\alpha}M^{j]\beta}_{\,\tck \,\tck\beta}\Big)\ ,\\
\label{459}
&&\tau_{\alpha\beta}^{\tck\tck i}=
\mathop{\tau_{\alpha\beta}^{\tck\tck i}}
\limits^{\textrm{conv}}+\partial_jM^{ij}_{\tck\tck\alpha\beta}+
M^i_{\tck[\alpha\beta]}+e^i_{\tck[\beta}M_{\alpha]\gamma}
^{\tck \,\tck\gamma}\ .
\eea
The convective currents are of the usual Schr$\ddot{\textrm o}$dinger
type
\bea
\label{460}
&&\mathop{\Sigma_\alpha^{\tck i}}
\limits^{\textrm{conv}}: =\frac{1}{2m}(\partial^i\bar\psi)
\partial_\alpha\psi+ h.c. +\delta^i_\alpha\, \mathop{{\cal L}}
\limits^{\textrm{conv}}\ ,             \\[4pt]
\label{461}
&&\mathop{\tau_{\alpha\beta}^{\tck\tck i}}
\limits^{\textrm{conv}}: =-\frac{1}{2m}
(\partial^i\bar\psi)
f_{\alpha\beta}\psi+ h.c.\ ,
\eea
with
\be
\label{462}
\mathop{{\cal L}}\limits^{\textrm{conv}}: =
-\frac{1}{2m}\Big[(\partial_\alpha\bar\psi)
\partial^\alpha\psi-m^2\bar\psi\psi\Big]\ .
\ee
In analogy with the Dirac-Maxwell theory (i.e. with Dirac plus
U(l)-gauge, whereas we have Dirac plus PG) in (\ref{458}),
(\ref{459}) there
emerge the translational and the rotational
\emph{gravitational moments} of the electron field
\be
\label{463}
M^{ij}_{\tck\tck\alpha}: =\frac{1}{m}\bar\psi f^{ji}
\partial_\alpha\psi
\ee
and
\be
\label{464}
M^{ij}_{\tck\tck\alpha\beta}: =\frac{1}{m}\bar\psi f^{ji}
f_{\alpha\beta}\psi\ ,
\ee
respectively. We stress that these two new expressions for the
gravitational moments of the electron are measurable in principle.
Hence there is a way to decide, whether it makes physical sense to
Gordon-decompose the momentum and the spin currents of the electron
field.

If we introduce the polarization currents
\be
\label{465}
\mathop{\Sigma_\alpha^{\tck i}}\limits^{\textrm{pol}}: =
\Sigma_\alpha^{\tck i}-\mathop{\Sigma_\alpha^{\tck i}}
\limits^{\textrm{conv}},\qquad
\mathop{\tau_{\alpha\beta}^{\tck\tck i}}
\limits^{\textrm{pol}}:\ = \tau_{\alpha\beta}^{\tck\tck i}-
\mathop{\tau_{\alpha\beta}^{\tck\tck i}}
\limits^{\textrm{conv}}\ ,
\ee
then we have finally
\bea
\label{466}
&&\mathop{\Sigma_\alpha^{\tck i}}\limits^{\textrm{pol}}=
\partial_j\Big( M^{ij}_{\tck\tck\alpha}+2
e^{[i}_{\,\tck\alpha}M^{j]\gamma}_{\tck \ \tck\gamma}\Big)\ ,\\
\label{467}
&&\mathop{\tau_{\alpha\beta}^{\tck\tck i}}
\limits^{\textrm{pol}}=\partial_jM^{ij}_{\tck\tck\alpha\beta}
+M^i_{\tck[\alpha\beta]}+e^i_{\tck[\beta}M_{\alpha]\gamma}
^{\tck \ \tck\gamma}\ .
\eea
Observe that the last two terms in (\ref{467}) are required by angular
momentum conservation (\ref{313}).

Up to now we just carried through some special-relativistic
Dirac algebra. Let us now turn to our field equations (\ref{344}),
(\ref{345}) and linearize them:
\bea
\label{468}
&&\Sigma_\alpha^{\tck i}\sim\partial{\cal H}_\alpha^{\tck ij}\ ,\\[3pt]
\label{469}
&&\tau_{\alpha\beta}^{\tck\tck i}\sim\partial{\cal H}_
{\alpha\beta}^{\,\tck\tck ij}+
{\cal H}_{[\alpha\beta]}^{\,\tck\tck \ i}
\eea
A comparison with (\ref{466}), (\ref{467}) will reveal
immediately a similarity in structure. Read off the working hypothesis:
The translational and rotational field strengths couple
in an analogous way to the canonical currents, as the respective
gravitational moments to the polarization currents. Consequently we get
\bea
\label{470}
&&{\cal H}_\alpha^{\tck ij}\sim F^{ij}_{\tck\tck\alpha}
e^{[i}_{ \ \tck\alpha}F^{j]\gamma}_{\tck \ \tck\gamma}\,\\[3pt]
\label{471}
&&{\cal H}_{\alpha\beta}^{\,\tck\tck ij}\sim F^{ij}_{\tck\tck\alpha\beta}\ .
\eea
Properly adjusting the dimensions, leads straightaway to the field
lagrangian of our choice
\be
\label{472}
V =(e/4\ell^2)\Big(
-F^{ij}_{\tck\tck\alpha}F_{ij}^{\tck\tck\alpha}+2
F^{i\gamma}_{\tck\tck\gamma}F_{i\delta}^{\tck\tck\delta}\Big)
+(e/4\kappa)\Big(
-F_{ij\alpha}^{\tck\tck\tck\beta}F^{ij\alpha}_{\tck\tck\tck\beta}
\Big)
\ee
without all the involved reasonings used earlier.\symbolfootnote[1]
{According to Wallner \cite{69}, one can reformulate the
Gordon decomposition
such that one arrives at the choice $\lambda=0$. We are not able
to understand his arguments, however.}

The most remarkable achievement of the working hypothesis as
formulated above is the following: Without having ever talked
about gravity, one arrives at a lagrangian which, in the enforced
$T_4$-limit, yields the Schwarz\-schild solution, the newtonian
approximation, etc. Some consequences of (\ref{472}),
like a ``confinement" potential in a weak field approximation,
have been worked out already \cite{58,13}. But since we are
running out of space and time, we shall discuss these matters elsewhere.

\section*{Acknowledgements}
\addcontentsline{toc}{section}{Acknowledgements}

For many enlightening and helpful discussions on the PG I am
most grateful to Yuval Ne'eman as well as to J$\ddot{\textrm{u}}$rgen
Nitsch. I'd like to thank P. G. Bergmann and V. De Sabbata for inviting
me to give these lectures in Erice and I highly appreciate the support
of Y. Ne'eman and E. C. G. Sudarshan (CPT) and of J. A. Wheeler
(CTP) of The University of Texas at Austin, where these notes were
written up.



\addcontentsline{toc}{section}{Literature}
\begin{thebibliography}{99}
\setlength\itemsep{3pt}

\bibitem{1} F. W. Hehl, P. von der Heyde, G. D. Kerlick, and J. M. Nester,
\emph{Rev. Mod. Phys.} 48:393 (1976).

\bibitem{2} E. Cartan, \emph{C. R. Acad. Sci.} (Paris) 174:593 (1922).
A translation was provided by G. D. Kerlick, see his Ph.D. Thesis,
Princeton University (1975).

\bibitem{3} J. A. Schouten, Ricci Calculus, 2nd ed., Springer, Berlin
(1954).

\bibitem{4} Y. Choquet-Bruhat, C. DeWitt-Morette, and M. Dillard-Bleick,
``Analysis, Manifolds and Physics", North-Holland,
Amsterdam (1977).

\bibitem{5} L. O'Raifeartaigh, \emph{Rep. Progr. Phys.} 42:159 (1979).

\bibitem{6} H. Weyl, \emph{Z. Phys.} 56:330 (1929).

\bibitem{7} C. N. Yang and R. L. Mills, \emph{Phys. Rev.} 96:191 (1954).

\bibitem{8} R. Utiyama, \emph{Phys. Rev.} 101:1597 (1956).

\bibitem{9} D. W. Sciama, Recent Developments in General Relativity
(Festschrift for Infeld), Pergamon, Oxford (1962), p.415.

\bibitem{10} T. W. B. Kibble, \emph{J. Math. Phys.} 2:212 (1961).

\bibitem{11} Y. Ne'eman, ``Gravity, Groups, and Gauges," Einstein
Commemorative Volume (A. Held et al., eds.), Plenum Press, New York, to
be published (1980).

\bibitem{12} A. Trautman, ``Fiber Bundles, Gauge Fields, and Gravitation,"
ibid.

\bibitem{13} F. W. Hehl, J. Nitsch, and P. von der Heyde, ``Gravitation and
the Poincare Gauge Field Theory with Quadratic Lagrangian,"
ibid.

\bibitem{14} Y. Ne'eman, this volume.

\bibitem{15} A. Trautman, this volume.

\bibitem{16} J. Nitsch, this volume.

\bibitem{17} H. Rumpf, this volume.

\bibitem{18} W. Szczyrba, this volume.

\bibitem{19} M. A. Schweizer, this volume.

\bibitem{20} P. B. Yasskin, see his Ph.D. Thesis, University of Maryland
(1979).

\bibitem{21} V. N. Tunyak, \emph{Sov. Phys. J.} (Izv. VUZ. Fiz., USSR) 18:74,77
(1975); 19:599 (1976); 20:538,1337,1537 (1977).

\bibitem{22} Current Problems in Theoret. Physics (Festschrift for Ivanenko),
Mos\-cow Univ. (1976).

\bibitem{23} Z. Zuo, P. Huang, Y. Zhang, G. Li, Y. An, S. Chen, Y. Wu, Z.
He, L. Zhang, and Guo, \emph{Scientia Sinica} 22:628 (1979).

\bibitem{24} P. von der Heyde, \emph{Phys. Lett.} 58A:141 (1976);
\emph{Z. Naturf.} 31a:1725 (1976).

\bibitem{25} P. von der Heyde, talk given at the ``9th Texas Symposium on
Relativistic Astrophysics," during the Workshop ``Gauge Theories of
Gravity", held at Munich in December 1978 (unpublished), and
unpublished draft.

\bibitem{26} Y. Ne'eman, Lecture Notes in Mathematics (Springer)
No. \emph{676}, p.189 (1978).

\bibitem{27} R. Jackiw, \emph{Phys. Rev. Lett.} 41:1635 (1978).

\bibitem{28} F. Mansouri and C. Schaer, \emph{Phys. Lett.} 83B:327 (1979).

\bibitem{29} J. M. Nester, Ph.D. Thesis, University of Maryland (1977).

\bibitem{30} R. J. Petti, \emph{Gen. Rel. Gravitation J.} 7:869 (1976).

\bibitem{31} Y. Ne'eman and T. Regge, \emph{Riv. Nuovo Cimento} 1,
No. 5, 1 (1978).

\bibitem{32} G. Cognola, R. Soldati, M. Toller, L. Vanzo, and S. Zerbini,
Preprint Univ. Trento (1979).

\bibitem{33} Y. Ne'eman and E. Takasugi, Preprint University of Texas at
Austin (1979).

\bibitem{34} P. von der Heyde, \emph{Lett. Nuovo Cim.} 14:250 (1975).

\bibitem{35} H. Meyer, Diploma Thesis, University of Cologne (1979).

\bibitem{36} P. von der Heyde and F. W. Hehl, Proc. of the First Marcel
Grossmann Meeting on General Relativity, held at Trieste
in 1975 (R. Ruffini, ed.) North-Holland, Amsterdam (1977), p.255.

\bibitem{37} S. Hojman, M. Rosenbaum, M. P. Ryan, and L. C. Shepley,
\emph{Phys. Rev.} D17:3141 (1978).

\bibitem{38} S. Hojman, M. Rosenbaum, and M. P. Ryan,
\emph{Phys. Rev.} D19:430 (1979).

\bibitem{39} W.-T. Ni, \emph{Phys. Rev.} D19:2260 (1979).

\bibitem{40} C. Mukku and W. A. Sayed, \emph{Phys. Lett.} 82B:382 (1979).

\bibitem{41} F. W. Hehl, \emph{Rep. on Math. Phys.} (Torun) 9:55 (1976).

\bibitem{42} Y. M. Cho, \emph{J. Phys.} A11:2385 (1978).

\bibitem{43} S. Hojman, \emph{Phys. Rev.} 18:2741 (1978).

\bibitem{44} A. P. Balachandran, G. Marmo, B. S. Skagerstam, and A. Stern,
CERN-preprint TH.2740 (1979).

\bibitem{45} W. R. Stoeger and P. B. Yasskin, \emph{Gen. Rel. Gravitation J.}
(to be published).

\bibitem{46} P. B. Yasskin and W. R. Stoeger, Preprint Harvard University
(1979).

\bibitem{47} E. J. Post, ``Formal Structure of Electromagnetics", North-
Holland, Amsterdam (1962).

\bibitem{48} H. Rund, \emph{J. Math. Phys.} 20:1392 (1979).

\bibitem{49} C. W. Misner, K. S. Thorne, and J. A. Wheeler, ``Gravitation",
Freeman, San Francisco (1973).

\bibitem{50} P. van Nieuwenhuizen, this volume.

\bibitem{51} J. Wess, talk given at the Einstein Symposium Berlin, preprint
Univ. Karlsruhe (1979).

\bibitem{52} F. W. Hehl, G. D. Kerlick, and P. von der Heyde,
\emph{Z. Naturforsch.} 31a:111,524,823 (1976); \emph{Phys. Lett.}
63B:446 (1976).

\bibitem{53} F. W. Hehl, E. A. Lord, and Y. Ne'eman,
\emph{Phys. Lett.} 71B:432 (1977); \emph{Phys. Rev.} D17:428 (1978).

\bibitem{54} E. A. Lord, \emph{Phys. Lett.} 65A:1 (1978).

\bibitem{55} Y. Ne'eman and Dj. \v Sija\v cki,
\emph{Ann. Phys.} (N.Y.) 120:292 (1979).

\bibitem{56} Y. Ne'eman, in ``Jerusalem Einstein Centennial Symposium",
(Y. Ne'eman, ed.), Addison-Wesley, Reading, to be published
(1980).

\bibitem{57} S. J. Aldersley, \emph{Gen. Rel. Gravitation J.} 8:397 (1977).
\bibitem{58} F. W. Hehl, Y. Ne'eman, J. Nitsch, and P. von der Heyde,
\emph{Phys. Lett.} 78B:102 (1978).

\bibitem{59} S. Hamamoto, \emph{Progr. Theor. Physics} 61:326 (1979).

\bibitem{60} V. De Sabbata and M. Gasperini,
\emph{Lett. Nuovo Cimento} 21:328 (1978)

\bibitem{61} K. Hayashi and T. Shirafuji, \emph{Phys. Rev.} D19:3524 (1979).

\bibitem{62} D.-E. Liebscher, this volume.

\bibitem{63} C. M\o ller, \emph{Mat. Fys. Medd. Dan. Vid. Selsk. 39},
No. 13 (1978).

\bibitem{64} J. Nitsch and F. W. Hehl, Preprint Univ. Cologne (1979).

\bibitem{65} J. M. Nester, \emph{Phys. Rev.} D16:2395 (1977).

\bibitem{66} K. S. Stelle and P. C. West, \emph{J. Phys.} A12:L205 (1979).

\bibitem{67} K. S. Stelle and P. C. West, Preprint Imperial College London
(1979).

\bibitem{68} H. Rumpf, \emph{Z. Natur forsch.} 33a:1224 (1978).

\bibitem{69} R. P. Wallner, Preprint Univ. Vienna (1979); \emph{Gen. Rel.
Gravitation J.} (to be published).

\bibitem{70} J. Anandan, Preprint Univ. of Maryland (1978);
\emph{Nuovo Cimento} (to be published).

\bibitem{71} J. S. Anandan, in ``Quantum Theory and Gravitation", Proc. New
Orleans Conf., May 1979 (A. R. Marlow, ed.), Academic Press,
New York (to be published).

\bibitem{72} E. E. Fairchild, \emph{Phys. Rev.} D16:2438 (1977).

\bibitem{73} F. Mansouri and L. N. Chang, \emph{Phys. Rev.} D13:3192 (1976).

\bibitem{74} D. E. Neville, \emph{Phys. Rev.} D18:3535 (1978).

\bibitem{75} D. E. Neville, Preprint Temple Univ. (1979).

\bibitem{76} S. Ramaswamy and P. B. Yasskin, \emph{Phys. Rev.} D19:2264 (1979).

\bibitem{77} F. W. Hehl and B. K. Datta, \emph{J. Math. Phys.} 12:1334 (1971).


\end{thebibliography}
\end{document}